\documentclass[aps,showpacs,floatfix]{revtex4}

\usepackage{graphicx}%
\usepackage{dcolumn,float}
\newcommand{\no}{\noindent}

\newcommand{\beq}{\begin{equation}}
\newcommand{\eeq}{\end{equation}}
\newcommand{\qq}{\begin{eqnarray}}
\newcommand{\qqq}{\end{eqnarray}}

\newcommand{\ve}{\varepsilon}
\begin{document}

\title{Large $N$  limit of the exactly solvable
BCS model: 
 \\ analytics versus numerics
}

\date{\today}

\author{J.M.\ Rom\'an$^{1}$, G.\ Sierra$^{1}$, and
 J.\ Dukelsky$^{2}$ } 

\affiliation{ 
$^{1}$Instituto de F\'{\i}sica Te\'orica, CSIC/UAM,
Madrid, Spain. \\
$^{2}$Instituto de Estructura de la Materia, CSIC, Madrid, Spain
 }

\begin{abstract}
We have studied the numerical solutions of Richardson
equations of the BCS model in the limit of large number
of energy levels at half-filling, 
and compare them with the analytic results 
derived by Gaudin and Richardson, which in turn leads to 
the standard BCS solution. We focus on the 
location and density of the roots, the eigenvalues
of the conserved quantities, 
and the scaling properties of the total energy for the equally spaced
and the two-level models. 
\end{abstract}

%\pacs{71.10.Hf, 75.10.Jm, 75.30.Gw, 74.20.Mn}
% PACS, the Physics and Astronomy Classification Scheme.

\maketitle

\newcommand{\bb}{\boldsymbol{\beta}}
\newcommand{\ba}{\boldsymbol{\alpha}}

\section*{I) Introduction}
\label{sec1:level1}

One of the most studied models in Condensed Matter
Physics is the BCS model of superconductivity, 
proposed by Bardeen, Cooper, and Schrieffer in 1957 \cite{BCS}. 
The BCS model is usually formulated in the
grand canonical ensemble, which is appropiated
to systems with a macroscopic number of fermions. However
for small systems, such as nuclei or ultrasmall
metallic grains, one has to consider the canonical
ensemble where the BCS wave function is not adequate.

Fortunately enough, there is a 
simplified version of the BCS model
which is exactly solvable in the canonical ensemble. 
This occurs when the strengh of the pairing interaction between all
the energy levels is the same. The exact solution 
of the reduced BCS model was obtained by Richardson
in 1963 \cite{R1,R1bis,RS}, who also studied
its consequences and properties
in a series of papers in the 60's and 70's 
\cite{R-norm,R-roots,R-boson,R-limit}. 
The exactly solvable BCS model, in turn, is closely related
to the rational spin model of Gaudin, defined
in terms of a set of commuting Hamiltonians \cite{Gaudin,G-book}.

The integrability of the reduced BCS
Hamiltonian was proved in 1997 by 
Cambiaggio, Rivas, and Saraceno (CRS) \cite{CRS}, who constructed
a set of conserved quantities in involution,   
commuting with the BCS Hamiltonian and closely related to 
the rational Gaudin's Hamiltonians \cite{Gaudin}. 
More recently Gaudin's trigonometric 
model have been generalized, to include a $g$-term \`a la BCS,
in references \cite{ALO,DES}. There are also bosonic pairing Hamiltonians
satisfying the same type of 
equations as in the fermionic case \cite{R-boson,DS,DP}. 
Generalizations
of the $SU(2)$ BCS  model 
to other Lie groups \cite{AFS}
and supergroups \cite{KM}
have been worked out.
Both, the BCS and Gaudin's models are intimately linked
to conformal field theory \cite{B,BF,S}, 
Chern-Simons theory \cite{AFS},
and Integrable Models in Statistical Mechanics \cite{AFF,ZLMG,vDP} 
(see ref.~[\onlinecite{Scomo}] for a short review on these topics).
The exact solution has also 
been used to study the effect of level statistics~\cite{random}
and finite temperature \cite{Finite-T,FFM}
on the ultrasmall metallic grains.

An important property of the exact solution
is that the energy of the states and the
occupation numbers agree, to leading order
in the number of particles, 
with the BCS theory \cite{BCS}. 
This result was obtained by Gaudin \cite{G-book}  
and Richardson \cite{R-limit}, 
using an electrostatic analogy of the 
Richardson's equations, which have to be
solved for finding the eigenstates of the BCS Hamiltonian.
More recent applications of this electrostatic
model to nuclear pairing can be found in ref.~[\onlinecite{DEP}]. 

In the limit of large number of levels $N$, 
Richardson's equations become integral equations,
which can be solved with techniques 
of complex analysis \cite{G-book}.  
This is sufficient to determine 
the location and density of roots,
as well as the total energy of the ground state
to  leading order in $N$. To study higher
order corrections of the ground state energy and the excitations,
Richardson developed a multipole expansion of the 
fields appearing in the electrostatic analogue model \cite{R-limit}. 
We shall follow Gaudin's approach
which is more geometrical, but use also some of the results
obtained by Richardson.

The aim of this paper is to make a revision and 
precise comparison 
between the numerical solution and the analytic
expressions for the location and density of 
roots of the Richardson's equations, as proposed
originally by Gaudin in his unpublished paper
in 1968, which got finally published in his
collected papers in 1995 \cite{G-book}. 
The comparison will be 
made in the limit of large 
$N$ for  the two-level model and the equally
spaced model. 

These models have been extensively used in Nuclear Physics 
since they were introduced by Hogaasen-Feldman 
\cite{Ho} (two-level) and Richardson himself \cite{R-roots}
(equally spaced). The latter model has also been used to
describe the physics of ultrasmall metallic grains 
\cite{grains,Lanczos,BvD,DMRG,Sch} (for a review see ref.~[\onlinecite{vDR}]).
In the large $N$ limit 
the equally spaced model is superconducting
for all values of atractive BCS couplings $g$, 
while the two-level model displays a quantum 
phase transition between a superconducting state
and a normal state as a function of the BCS coupling constant.

The organization of the paper is as follows:
in sections II and III we give a primer
of the Richardson's exact solution of the BCS
model, together with its integrability.
 In section IV we introduce the electrostatic analogue model
of BCS. In section V we review Gaudin's continuum limit
of the Richardson's equations and their general solution,
using complex analysis. In section VI we derive
the gap and chemical potential BCS equations
under the assumption that the roots of Richardson
equations collapse into a single arc, and study in detail
the two-level and equally spaced models. In section
VII we present our numerical results and compare
them with the analytic ones obtained in section VI. We also study
the scaling properties of the roots and the ground
state energy. In section VIII we present our
conclusions and prospects. Finally, we describe in an appendix
the conformal mappings associated
to the equally spaced model.

\section*{II) Richardson's exact solution of the reduced
BCS model}

The reduced BCS model is defined by the Hamiltonian 
\cite{BCS,vDR,grains}

\beq
H_{BCS} = \frac{1}{2}\sum_{j, \sigma= \pm} 
\varepsilon_{j\sigma} c_{j \sigma}^\dagger c_{j \sigma}
  - G \sum_{j, j'}  c_{j +}^\dagger c_{j -}^\dagger 
c_{j' -} c_{j' +} \; , 
\label{1}
\eeq

\noindent where $c_{j,\pm}$ (resp.\ $c^\dagger_{j,\pm}$)
is an electron  
annihilation (resp.\ creation) operator   
in the time-reversed states $|j, \pm \rangle$
with energies $\varepsilon_j/2$, and
$G$ is the BCS dimensionful coupling constant.
The sums in (\ref{1}) run over a set of $N$ doubly
degenerate energy levels $\varepsilon_j/2 \; (j=1,\dots, N)$.
We adopt Gaudin's notation, according to which
$\varepsilon_j$ denotes the energy of a 
pair occupying the level $j$ \cite{G-book}. 
We shall assume that the energy levels  are all
distinct, i.e.\ $\varepsilon_i \neq \varepsilon_j $ for $i \neq j$.  
The Hamiltonian (\ref{1}) is a simplified 
version of the reduced BCS Hamiltonian where all couplings
have been  set equal to a single one,  namely $G$. 
Hereafter  we shall refer to  
(\ref{1}) simply as the BCS Hamiltonian. 

Richardson had long ago solved this model exactly for 
an arbitrary set of levels, $\varepsilon_j$ 
\cite{R1,R1bis,RS}. 
To simplify matters,  we shall assume that
there are not singly occupied electronic levels. 
As can be seen from (\ref{1}), these levels decouple
from the rest of the  system;  
they are said to be blocked,  
contributing only  with their energy  
$\varepsilon_j/2$ to the total energy $E$. 
The above simplification implies that every energy level
$j$ is either empty (i.e.\ $| {\rm vac}\rangle$),
or occupied by a pair of electrons (i.e.\
$c^\dagger_{j,+} c^\dagger_{j,-} |{\rm vac}\rangle$). 
Denote the total number of electrons pairs by 
$M$. Then  of course
$M \leq N$. The most studied case
in the literature corresponds to the half-filled situation,  
where the number of electrons,  $N_e= 2M$, equals the number of levels
$N$ \cite{vDR}. 
In the absence of interaction (i.e.\ $G=0$),  
all the  pairs occupy the lowest  energy levels, forming a  
Fermi sea. The pairing interaction promotes
the  pairs to  higher  energies and even
for small values of $M$ or $G$ the levels are pair correlated
\cite{Lanczos,BvD,DMRG,vDR}.

In order to describe Richardson's solution one
defines the hard-core boson operators

\beq
b_j = c_{j,-} c_{j,+}, \qquad 
b_j^\dagger= c^\dagger_{j,+} c^\dagger_{j,-} , \qquad 
N_j = b^\dagger_j b_j \; ,
\label{2}
\eeq

\noindent which satisfy the commutation relations,

\beq
[ b_j, b_{j'}^\dagger ] = \delta_{j,j'} \; (1 - 2 N_j). 
\label{3}
\eeq

The Hamiltonian (\ref{1}) can then be written as

\beq
H_{BCS} = \sum_{j}  \varepsilon_j b^\dagger_j b_j - G \,
\sum_{j,j'} \; b_j^\dagger b_{j^{\prime}} \; .
\label{4}
\eeq

\noindent 
Richardson showed that the eigenstates of this Hamiltonian 
with $M$ pairs have the (unnormalized) product form 
\cite{R1,R1bis,RS}

\beq
|M \rangle = \prod_{\nu = 1}^M B_\nu |{\rm vac} \rangle, 
\qquad
B_\nu = \sum_{j=1}^N \frac{1}{\varepsilon_j - E_\nu} 
\; b^\dagger_j \; ,
\label{5} 
\eeq

\noindent 
where the  parameters $E_\nu$ ($\nu = 1, \dots , M$) are, 
in general, complex solutions of the $M$ coupled algebraic
equations 

\beq
\frac{1}{G } 
= \sum_{j=1}^N \frac{1}{ \varepsilon_j - E_\nu}
- \sum_{\mu=1 (\neq \nu)}^{M} \frac{2}{ E_\mu - E_\nu}\;, 
\qquad \nu =1, \dots, M 
\label{6}
\eeq

\noindent 
which play the role of Bethe ansatz equations for this
problem \cite{B,BF,S,AFF,ZLMG,vDP,Scomo}. 
The energy of these states is given by the sum of the
auxiliary parameters $E_\nu$, i.e.\

\beq
{E} (M) = \sum_{\nu = 1}^{M} E_\nu.
\label{7}
\eeq

\noindent 
The ground state  of $H_{BCS}$ is given by the solution
of eqs.~(\ref{6}) which gives the lowest value of ${E}(M)$. 
The (normalized) states (\ref{5}) can also be written  
as \cite{R-norm}

\beq
|M \rangle = \frac{C}{ \sqrt{M!} } \sum_{j_1, \cdots , j_{M}}
\psi(j_1, \dots,  j_{M}) b^\dagger_{j_1} \cdots b^\dagger_{j_{M}}
|{\rm vac} \rangle, 
\label{8}
\eeq

\noindent where 
the sum excludes double occupancy of pair states
and the wave function $\psi$ takes the form

\beq
\psi(j_1, \cdots, j_{M}) = \sum_{\cal P}
 \prod_{k=1}^{M} \frac{1}{ \varepsilon_{j_k} 
- E_{{\cal P}k} }\; .  
\label{9} 
\eeq

\noindent The  sum in (\ref{9})  runs  over all the 
permutations, ${\cal P}$, 
of $1, \cdots, M$. The constant  $C$ in (\ref{8})
guarantees the normalization of the state \cite{R-norm} (i.e.\
$\langle M|M\rangle =1$).

\section*{III) Integrability of the reduced BCS Hamiltonian}

A well known fact about the BCS Hamiltonian is that
it is  equivalent to that of a XY spin model with long
range couplings, and a ``position dependent'' 
magnetic field proportional to $\varepsilon_j$.
To see this, let us represent the hard-core boson operators
(\ref{2})  in terms of the Pauli 
matrices as follows,

\beq
b_j = \sigma_j^+, \qquad 
b_j^\dagger = \sigma_j^-, \qquad 
N_j = \frac{1}{2} (1 - \sigma^z_j), 
\label{10}
\eeq

\noindent 
in which case  the Hamiltonian (\ref{4}) becomes

\begin{eqnarray}
H_{BCS} & = & H_{XY} + \frac{1}{2}\sum_j \varepsilon_j +
 G (N/2 -M),  \label{11} \\
H_{XY} & = & - \sum_j  \varepsilon_j t^0_j - \frac{G}{2}
(T^+ \; T^- + T^- \; T^+ ),  \nonumber 
\end{eqnarray}

\noindent where the matrices

\beq
T^a = \sum_{j=1}^N  t^a_j, \qquad (a = 0, +, -),
\qquad \mbox{with} \qquad
t^0_j = \frac{1}{2} \sigma^z_j, \quad 
t^+_j = \sigma^+_j, \quad 
t^-_j = \sigma^-_j,
\label{12} 
\eeq 

\noindent satisfy the $SU(2)$ algebra,

\beq
[T^a, T^b] = f^{a b}_c T^c, \qquad \mbox{with}\qquad f^{+0}_+ = f^{0-}_- = -1, \quad f^{+-}_0 = 2, \label{13}
\eeq

\noindent whose Casimir is given by

\beq
{\bf T} \cdot {\bf T} = T^0 T^0 +
\frac{1}{2} (T^+ T^- + T^- T^+). 
\label{14}
\eeq

This spin representation of the BCS model
is the appropiate one to 
study its integrability, as it was shown 
by Cambiaggio, Rivas, and Saraceno (CRS),
who constructed a set
of operators, $R_j \;\;(j=1, \dots,N)$ \cite{CRS},

\begin{equation}
R_i = - t^0_i - 2 g \;  \sum_{j (\neq i)}^N
\frac{ {\bf t}_i \cdot {\bf t}_j }{ \varepsilon_i- \varepsilon_j},
\qquad
(i=1, \dots, \Omega),
\label{15} 
\end{equation}

\noindent 
satisfying the properties

\beq
[H_{BCS}, R_i] = 
 [R_i, R_j] = 0,  \qquad(i, j =1, \dots, N). \label{16} 
\eeq

\noindent 
Let us denote by $\lambda_j$ the eigenvalue of $R_j$ acting on the state 
(\ref{8}), i.e.,

\begin{equation}
R_j |M \rangle_R = \lambda_j |M \rangle_R.
\label{17}
\end{equation}
 
\noindent 
CRS, unaware of Richardson's  
solution,  did not give  
an expression of $\lambda_j$ in their work. 
This was found in reference [\onlinecite{S}]
using CFT techniques,

\begin{equation}
\lambda_i = - \frac{1}{2} + 
G \left(\sum_{\nu=1}^M \frac{1}{ \varepsilon_i - E_\nu}
- \frac{1}{2} \sum_{j=1 (\neq i)}^{N}  \frac{1}{ 
\varepsilon_i - \varepsilon_j} \right), 
\label{18}
\end{equation}

\no
which verifies $\sum_i \lambda_i = (M - N/2)$ upon using eqs.~(\ref{6}), 
so that it vanishes at half filling.

However CRS  showed  that  $H_{XY}$ given in
eq.~(\ref{11}) can be expressed in terms of the operators
$R_i$ as

\begin{equation}
H_{XY} = \sum_j  \varepsilon_j R_j + G \left(\sum_j R_j \right)^2 
- \frac{3}{4} G N. 
\label{19}
\end{equation}

\no 
Then using the eqs.~(\ref{11}), (\ref{18}), and (\ref{6})
one can show that ${E}(M)$ is indeed given by eq.~(\ref{7}).
(For the two-level system the last term in (\ref{19}) must be replaced 
by $-2G (N/4 +1)N/4$ ).

\section*{IV) The Gaudin-Richardson's electrostatic model of BCS}

The equations (\ref{6}) admit a 2 dimensional
electrostatic analogy, which will be very useful in the study 
of the limit $N\rightarrow \infty$ \cite{G-book,R-limit}. 
Let us consider a set of $N$ charges $-1/2$
fixed at the positions
$\varepsilon_j$ on the real axis, and a uniform field parallel
to this axis with stregth $- 1/2G$. The problem is to find
the equilibrium positions of $M$ charges $+1$ at positions 
$E_\mu$ subject to their mutual repulsion, the attraction with
the $-1/2$ charges, and the action of the uniform field.
The 2D electrostatic potential 
is given by $W + W^*$, where $W$ reads,

\begin{eqnarray}
W(\{E_\mu\},\{\varepsilon_i \}) & = &
\frac{1}{2} \sum_{i, \mu} {\rm log} (\varepsilon_i - E_\mu)
-   \sum_{\mu > \nu} {\rm log}
(E_\nu  - E_\mu)  \nonumber \\ 
& & - \frac{1}{4} \sum_{i > j} {\rm log} (\varepsilon_i - \varepsilon_j)
+ \frac{1}{2 G} \left(\sum_\mu E_\mu -  \frac{1}{2} 
\sum_i \varepsilon_i \right).
\label{20}
\qqq

\no
It is easy to see that the Richardson's eqs.~(\ref{6})
arise as the stationary conditions $\partial W/\partial E_\mu = 0$, 
while the conserved quantities (\ref{18}) 
are proportional to the forces
exerted on the fixed charges, i.e.\

\beq
\lambda_i = 2 G \;  \frac{ \partial W}{\partial \varepsilon_i}.
\label{21}
\eeq

\no 
The total energy (\ref{7}) is the center of gravity
of the charges $E_\mu$, which must be located symmetrically
with respect to the real axis, i.e.\ if $E_\mu$ is a solution
to the eqs.~(\ref{6}),  
then $E^*_\mu$ must also be a solution. This condition is fullfilled
in two cases: i) either $E_\mu$ is real, or ii) $E_\mu$ and  $E^*_\mu$
form a complex conjugate pair. 
The formation of these complex pairs is usually 
an indication of the superconducting properties
of the ground state.

\section*{V) Continuum limit of Richardson's equations}

In reference [\onlinecite{G-book}] Gaudin proposed a continuum limit
of the equations (\ref{6}), in order to obtain
the BCS formulas for the ground state energy and occupation
number of levels. 
We shall follow closely this reference.  
This continuum limit is defined by taking the number
of levels $N$ going to infinity, while keeping fixed
the following quantities:

\beq
g = G N , \qquad M/N.
\label{22}
\eeq

\no
At the same time the pair energy levels $\ve_i$
will be equivalent to a negative charge density
$- \rho(\ve)$ located on an interval $\Omega$
of the real axis. The total charge of this interval
is given by

\beq
- \int_\Omega \rho(\ve) d \ve = - \frac{N}{2}\; .
\label{23}
\eeq

\no We shall suppose that $\Omega$ is
contained inside an interval $(- \omega, \omega)$.
For example, in the BCS model 
with equally spaced levels, 
$\omega/2$ is equal to the Debye energy.

The basic asumption made by Gaudin,  which is supported
by numerical results, is that the solutions of eqs.~(\ref{6})
organize themselves into arcs $\Gamma_k \; (k=1, \dots, K)$,  
which are piece-wise differentiable and  symmetric
under reflection on the real axis. We shall call
$\Gamma$ the union of all these arcs, and $r(\xi)$
the linear charge density of roots $E_\mu$
in the complex plane. Hence, the total number
of pairs, $M$, and the total energy, $E$,  are given by

\qq
\int_\Gamma r(\xi) \; |d \xi| & = & M,  \label{24} \\
\int_\Gamma \xi \; r(\xi) |d \xi| & = & E. \label{25} 
\qqq

The continuum limit of eqs.~(\ref{6}) is

\beq
\int_\Omega \frac{ \rho(\ve) \; d \ve}{\ve - \xi}
- P \int_{\Gamma} \frac{ r(\xi') \; |d \xi'|}{ \xi' - \xi}
- \frac{1}{2 G} = 0, 
\qquad \xi\in \Gamma,
\label{26}
\eeq

\no
which implies that the total electric field
on every point of the arcs $\Gamma_k$ is null. 
On the other hand, the values of the conserved
quantities (\ref{18}) are given in the continuum by

\beq
\lambda(\ve) = - \frac{1}{2} +
G \left(\int_\Gamma \frac{ r(\xi) \; |d \xi|}{
\ve - \xi} - P \int_\Omega \frac{ \rho(\ve') \; d \ve'}{
\ve - \ve'} \right)\; , \qquad \ve \in \Omega. 
\label{27}
\eeq

The formal solution of eqs.~(\ref{26}) can be found as follows.
First of all, let us orient each arc $\Gamma_k$ from the point
$a_k$ to the point $b_k$, and call $L_k$ an anticlockwise
path encircling $\Gamma_k$. We look for an analytic field
$h(\xi)$ outside $\Gamma$ and the set $\Omega$, such that

\beq
r(\xi) \; |d \xi| = \frac{1}{2 \pi i}
(h_+(\xi) - h_-(\xi) ) \; d \xi, 
\qquad \xi \in \Gamma,
\label{28}
\eeq

\no
where $ h_+(\xi)$ and $ h_-(\xi)$ denote the limit
values of $h(\xi)$ to the right and left of $\Gamma$.
This can be understood using the electrostatic equivalence,
considering that the electric field presents a discontinuity proportional
to the superficial density of charge when crossing such surface.
Next, we define a function $R(\xi)$, 
with cuts along the curves $\Gamma_k$, by the equation

\beq
R(\xi) = \left[ \prod_{k=1}^K (\xi - a_k) (\xi - b_k)
\right]^{1/2}, 
\label{29}
\eeq

\no
and look for a solution which vanishes at the
boundary points of $\Gamma$, in the form

\beq
h(\xi) = R(\xi) \; \int_\Omega
 \frac{ \varphi(\ve) \; d \ve }{ \ve - \xi}\; .
\label{30}
\eeq

\no This field has to be constant at infinity,
hence the first $K-2$ moments of $\varphi$
must vanish, i.e.\

\beq
\int_\Omega \; \ve^\ell \; \varphi(\ve) = 0,
 \qquad 0  \leq \ell < K-1.
\label{31}
\eeq

\no This formula is empty if $K=1$. 
The contour integral surrounding the charge
density $r(\xi)$ in (\ref{26}) can be expressed
as

\beq
\int_\Gamma \frac{ r(\xi') \; |d \xi'|}{
\xi - \xi'} = \frac{1}{2 \pi i}  \int_\Gamma
\frac{ h_+(\xi') - h_-(\xi')}{
\xi - \xi'} \; d\xi' = \frac{1}{2 \pi i} \int_L 
\frac{ h(\xi')}{
\xi - \xi'} \; d\xi', 
\qquad \xi \in {\bf C} \Gamma,
\label{32}
\eeq

\no where $ {\bf C} \Gamma$ is the region
outside the curves in $\Gamma$. Using 
eqs.~(\ref{29}) and (\ref{30}), one finds
for the principal value  of (\ref{32})

\beq
P  \int_\Gamma \frac{ r(\xi') \; |d \xi'|}{
\xi - \xi'} = 
\frac{1}{2 \pi i} \int_L
\frac{ d\xi' R(\xi')}{ \xi - \xi'}  
\int_\Omega \frac{ \varphi(\ve) \; d \ve}{
\ve - \xi'} 
= - \int_\Omega \frac{ \varphi(\ve) R(\ve)}{
\ve - \xi}  d\ve + 
\int_\Omega \ve^{K-1} \varphi(\ve)  d\ve, 
\qquad \xi \in \Gamma,
\label{33}
\eeq

\no where we have deformed the contour of integration
$L$ into two contours, one encircling the 
interval $\Omega$ (first term) and another
one around the infinity (second term).
We are assuming that $\Gamma$, and consequently
$L$,  do not cut the interval $\Omega$. 
The case where $L$ intersects $\Omega$ actually
arises in the equally spaced model, and will
be discussed later on.

Plugging eq.~(\ref{33}) into (\ref{26}),
we see that a solution is obtained provided
the following equations are satisfied:

\qq
& &\varphi(\ve) = \frac{\rho(\ve)}{R(\ve)},
\label{34} \\
& & \nonumber \\
& &\int_\Omega \ve^{K-1} \frac{\rho(\ve)}{R(\ve)}
d \ve = \frac{1}{2 G}\; ,
\label{35}
\qqq

\no
which have to be suplemented by eq.~(\ref{31}),
which can be rewritten as

\beq
\int_\Omega \ve^{\ell} \frac{\rho(\ve)}{R(\ve)}
d \ve = 0, 
\qquad 0 \leq \ell < K-1 .
\label{36}
\eeq

\no The field $h(\xi)$ gives also the density
of charges $r(\xi)$, 

\beq
r(\xi) = \frac{1}{\pi} |h(\xi)|\; , 
\qquad \xi \in \Gamma,
\label{37}
\eeq

\no
and its value is given by replacing 
(\ref{34}) into (\ref{30}), i.e.\

\beq
h(\xi) = R(\xi) \; \int_\Omega
\frac{ \rho(\ve)}{R(\ve)} \; \frac{ d \ve }{ \ve - \xi}\; ,
\label{38}
\eeq

\no 
whose value at infinity is $- 1/2G$.  
Moreover, using eq.(\ref{32}) and performing the same
contour deformations that lead to eq.(\ref{33}),   
one can show that the conserved quantities
(\ref{27}) are given by

\beq
\lambda(\ve) = P\left[G R(\ve) \; \int_\Omega
\frac{ \rho(\ve')}{R(\ve')} \; \frac{ d \ve' }{ \ve' - \ve}
\right] =
\frac{G}{2}\left(h(\ve + i\; 0 ) + h(\ve - i\; 0) \right) ,
\qquad \ve \in \Omega, 
\label{39}
\eeq

\no
so that $\lambda(\ve)$ is the principal
part of $h(\ve)$ on the set $\Omega$. 
Finally the equations fixing the arcs $\Gamma_k$ 
are the equipotential curves of the total
distribution, i.e.\

\beq
{\cal R} \int_{a_k}^\xi h(\xi') \; d \xi' = 0, 
\qquad \xi \in \Gamma_k. 
\label{40}
\eeq

\section*{VI) The BCS equations}

In this section we shall show how the BCS
equations describing the ground state of the model
can be derived from the formalism of section V \cite{G-book}. 

The basic assumption is that for 
the ground state all the roots $E_\mu$ form a single
arc, i.e.\ $K=1$, leaving only
two complex parameters $a \equiv a_1$
and $b \equiv b_1$, which shall be denoted as

\beq
a = \ve_0 - i \Delta,\qquad b = \ve_0 + i \Delta,
\label{41}
\eeq

\no where $\ve_0$ and $\Delta >0$ are real 
parameters.

 From eq.~(\ref{38}) the electrostatic
field is given by

\begin{equation}
h(\xi) = 
 [(\xi - \ve_0)^2 + \Delta^2 ]^{\frac{1}{2}} 
\int_\Omega
 \frac{ \rho(\ve)}{  [ (\ve - \ve_0)^2 + \Delta^2 ]^{\frac{1}{2}} 
}  \frac{ d \ve }{ \ve - \xi}\; .
\label{42} 
\end{equation}

\no
And the conserved quantities $\lambda(\ve)$ can be derived
from the eq.~(\ref{39}).
Eq.~(\ref{35}) yields the BCS gap equation

\beq
\int_\Omega  \frac{\rho(\ve) \; d\ve}{
 \sqrt{(\ve - \ve_0)^2 + \Delta^2}}
 = \frac{1}{2 G}\; .
\label{43}
\eeq

\no Similarly, eq.~(\ref{24}) becomes
the chemical potential equation

\begin{equation}
M = \frac{1}{2 \pi i} \int_L h(\xi)  d \xi
= \int_\Omega \left( 
1 - \frac{ \ve - \ve_0}{ 
 \sqrt{(\ve - \ve_0)^2 + \Delta^2}}
\right) \rho(\ve)  d \ve,
\label{44}
\end{equation}

\no
while eq.~(\ref{25}) gives the BCS expression
for the ground state energy,

\beq
 E  = \frac{1}{2 \pi i} \int_L \xi\;  h(\xi) \; d \xi  
= - \frac{\Delta^2}{ 4 G} + 
\int_\Omega \left( 
1 - \frac{ \ve - \ve_0}{ 
 \sqrt{(\ve - \ve_0)^2 + \Delta^2}}
\right) \rho(\ve) \; \ve \;  d \ve. 
\label{45}
\eeq

\no  Gaudin's paper \cite{G-book} contains a
misprint in this equation since  
the term $- \frac{\Delta^2}{ 4 G} $
is quoted as $- \frac{\Delta^2}{ 2 G}$.

Comparing these equations with the 
corresponding ones in the BCS theory,  
we deduce the following relations
between  $\Delta$, $\ve_0$, and $\rho(\ve)$,   
and $\Delta_{BCS}$ (BCS gap),  
$\mu$ (chemical potential), and
$n(\epsilon)$ (single particle energy density):

\beq
\Delta = 2 \Delta_{BCS}, \qquad 
\ve_0 = 2 \mu, \qquad
\rho(\ve) = \frac{1}{4}\; n\left( \frac{\ve}{2} \right). 
\label{46}
\eeq

\no The factor $1/4$, in the last term
of (\ref{46}),
emerges from the normalization $1/2$
of the charge density
$\rho(\ve)$, times another factor $1/2$, 
due to the fact that 
the separation between energy pairs
is twice the separation between single particle
energy levels.

To determine the equation of the curve $\Gamma$
we define the complex variable

\beq
z = \sqrt{ (\xi - \ve_0)^2 + \Delta^2} = x + i y\;.
\label{47}
\eeq

\no Using eqs.~(\ref{40}) and (\ref{42})
we deduce for $\Gamma$,

\beq
0 = \int_\Omega d \ve \; \rho(\ve) 
\left\{ 
\frac{x}{  \sqrt{ (\ve- \ve_0)^2 + \Delta^2} }
+ \frac{1}{4} {\rm log} 
\frac{
\left( x -  \sqrt{ (\ve- \ve_0)^2 + \Delta^2} \right)^2 + y^2}{
\left( x +  \sqrt{ (\ve- \ve_0)^2 + \Delta^2} \right)^2 + y^2}
\right\}.
\label{48}
\eeq

\no Eqs.~(\ref{43}), (\ref{44}), and (\ref{45})
have been derived under the condition that
$\Gamma$ does not cut the set $\Omega$, which
requires that (\ref{48}) has no 
solutions with $\xi \in \Omega$. 
In the equally spaced model the curve 
$\Gamma$ may intersect $\Omega$, so this case
has to be treated with some care (see below). 

Finally, the mean occupation  
of the energy level
$\ve_j$ can be computed from the formula
$\langle N_j \rangle= 
\partial E/\partial \ve_j$. In the large
$N$ limit $\langle N_j \rangle$ 
can be shown to be given by \cite{G-book,R-limit}

\beq
N(\ve) = \frac{1}{2} \left( 1 - 
\frac{ \ve - \ve_0}{ \sqrt{ (\ve - \ve_0)^2 + \Delta^2}}
\right), \qquad \ve \in \Omega,
\label{ocupa}
\eeq 

\no which agrees with the BCS result \cite{BCS}. 
Next we shall apply the previous formalism
to two models.

\subsection*{The two-level model}

Let us suppose that we have 
two energy levels $\pm \ve_1$, 
which can accomodate at most $N/2$ pairs each. 
The charge density is given by

\beq
\rho(\ve) = \frac{N}{4}
[\delta(\ve + \ve_1) + \delta(\ve - \ve_1)].
 \label{49}
\eeq

\no From eq.~(\ref{44}) we deduce that the
chemical potential vanishes, i.e.\
$\ve_0 =0$, while the gap eq.~(\ref{43}) 
gives

\beq
\ve_1^2 + \Delta^2 = g^2, \qquad 
g = G N.
\label{50}
\eeq

\no
This equation implies that $g$ must be  
greater than $\ve_1$.
The total energy of this solution follows
from (\ref{45}),

\beq
E = - \frac{N}{4 g} ( \ve_1^2 + g^2).
\label{53}
\eeq

\no
Integrating eq.~(\ref{42}) we obtain the electric field,
and accordingly the value of the conserved quantities

\beq
h(\xi) =  \frac{N\xi}{2g} 
           \frac{\sqrt{\xi^2 + \Delta^2}}{\ve_1^2 - \xi^2}
\quad
\longrightarrow
\quad
\lambda(-\ve_1) = \frac{1}{8g\ve_1}(2\ve_1^2 + g^2).
\label{53a}
\eeq

\no
Similarly one finds  $\lambda(\ve_1) = -\lambda(-\ve_1)$,
in agreement with the condition $ \sum_i \lambda_i = 0$
derived in sect.~III.
The equation of the curve (\ref{48}) becomes
in this case

\beq
\frac{4 x}{g} + {\rm log} 
\frac{ (x-g)^2 + y^2}{ (x+g)^2 + y^2} = 0,
\label{51}
\eeq

\no or equivalently 

\beq
x^2 + y^2 + g^2 = \frac{ 2 x g}{ {\rm th} ( 2 x/g)}\;,
\label{52}
\eeq

\no
where $x$ and $y$ are given by eq.~(\ref{47}).
In Gaudin's paper \cite{G-book} 
there is a misprint in  
eqs.~(\ref{53}) and  (\ref{52}).

From the numerical results shown in 
fig.~\ref{fig4}, when $g > \ve_1$
$\Gamma$ is an open arc between the imaginary
points $\pm i \Delta$ in the complex $\xi$-plane.
However when $g = \ve_1$, in the limit when
$N \rightarrow \infty$, the arc closes at the origin,
and for $g < \ve_1$ it remains closed surrounding 
the point $- \ve_1$. 

To treat this case Gaudin considers 
a general situation where $\Gamma$ is a closed curve 
surrounding a piece $\omega$ of $\Omega$
and defines a function $s(\xi)$ by the equation
 [\onlinecite{G-book}]

\beq
s(\xi) \; d \xi = r(\xi) \; |d \xi|.
\label{54}
\eeq

\no
Then eq.~(\ref{26}) is solved by

\beq
s(\xi ) = \frac{i}{\pi}
\left[  \frac{1}{2G} + 
\int_\omega \frac{ \rho(\ve) \;  d \ve}{
\ve - \xi}  
- \int_{ {\bf C}   \omega} \frac{ \rho(\ve) \;  d \ve}{
\ve - \xi} 
\right],
\label{55}
\eeq

\no which can be verified  using 
$P \int_{\Gamma} d \xi'/(\xi' - \xi) = i\pi$, when $\xi \in \Gamma$. 
For the two-level model $s(\xi)$ is given by

\beq
s(\xi) =  \frac{i}{\pi}
\left[  \frac{1}{2G} - 
\frac{N}{2} \frac{ \ve_1}{ \ve_1^2 - \xi^2} 
\right], 
\qquad \xi \in \Gamma.
\label{56}
\eeq

\no
and eqs.~(\ref{24}) and (\ref{25}) yield

\qq
M & = & \int_\Gamma d \xi \; s(\xi)  = \int_\omega
d \ve \;  2 \rho(\ve) = \frac{N}{2}\;,
\label{61} \\
& & \nonumber \\
E & = & \int_\Gamma d \xi \;  \xi   s(\xi)  = \int_\omega
d \ve \; 2 \ve \rho(\ve) = - \frac{N}{2} \ve_1\;, 
\label{62}
\qqq

\no Hence the state for $g < \ve_1$ 
is a Fermi sea to leading  order in $N$. 

The conserved quantities are derived  from an equation
similar to (\ref{39}), where $h(\xi)$ is replaced by $s(\xi)$, i.e.

\beq
\lambda(\ve) = -i\frac{\pi G}{2}
                 \left(s(\ve+i0) + s(\ve-i0)\right), \quad \ve \in \omega
\qquad \longrightarrow \qquad
\lambda(-\ve_1) = \frac{1}{2}\left(1-\frac{g}{4\ve_1}\right),
\label{62a}
\eeq

\no
verifing, as in the previous case, the symmetry 
$\lambda(\ve_1) = -\lambda(-\ve_1)$.

To find the curve $\Gamma$ we integrate (\ref{56}), 
and impose the result to be real accordingly to (\ref{54}).
Impossing the imaginary part of eq.~(\ref{54}) to vanish we 
obtain a diferential equation for the curve $\Gamma$, 
which is a check of consistency.

\beq
{\cal R} \left( 
\frac{ \xi - x_0}{ g} 
+ \frac{1}{2} \;  {\rm log} \;  \frac{ \xi - \ve_1}{ \xi + \ve_1}
\right) = 0, 
\qquad \xi \in \Gamma,
\label{57}
\eeq

\no where $x_0$ is an 
integration constant, which has  a non vanishing 
value, contrary to Gaudin's assumption.
Eq.~(\ref{57}) can be written as

\beq
x^2 + y^2 + \ve_1^2 = \frac{ 2  x  \ve_1}{ {\rm th} (2 (x - x_0)/g)},
\qquad \xi = x + i y.
\label{58}
\eeq

\no To find $x_0$ we have first to look for the 
point $\ve_0$ where $\Gamma$ intersects the real axis.
 From fig.~\ref{fig4}, the density of roots at this point 
vanishes, i.e.\ $s(\ve_0) = 0$, which implies

\beq
\ve_0 = - \ve_1 \;\sqrt{ 1 - g/\ve_1}\;.
\label{59}
\eeq

\no The value of $x_0$ is found by imposing
that $\ve_0$ belongs to the curve (\ref{58}), 
namely

\qq
x_0 = \ve_0 - \frac{g}{2} \; {\rm log} 
\frac{ \ve_1 + \ve_0}{ \ve_1 - \ve_0}\;.
\label{60}
\qqq

\no As $g$ approaches $\ve_1$ from below, 
both $\ve_0$ and $x_0$ go to zero, and the
curve (\ref{58}) coincides with the curve (\ref{52})
at the ``critical point'' $g = \ve_1$. 

\subsection*{Equally spaced model}

Let us consider now a model with $N$ pair
energy levels that are uniformly distributed in the
interval $[ -\omega, \omega]$, with a density 
$\rho(\ve) = \rho_0$

\beq
\rho_0 = \frac{N}{4 \omega}\; .
\label{63}
\eeq

\no The distance between the single particle
energy levels, $d$, is given by $d = \omega/N =
2 \omega_D/N$, where $\omega_D$ is the Debye energy.
This is the model used in most of the studies on ultrasmall
superconducting grains \cite{vDR}. At half-filling the number
of pairs is $M= N/2$, which implies that $\ve_0 =0$
(see eq.~(\ref{44})). On the other hand, the gap eq.~(\ref{43}) 
yields the standard result \cite{BCS}

\beq
\Delta = \frac{\omega}{{\rm sinh}(1/g)}, \qquad 
g = \frac{ G N}{\omega}\;,
\label{64}
\eeq

\no where $g$ is the usual dimensionless BCS coupling 
constant. Recall that $\Delta$ is twice the BCS gap
$\Delta_{BCS}$. The total energy of the BCS ground state
can be derived from (\ref{45}),

\beq
E = - \frac{\omega^2}{4 d} \; \sqrt{ 1 + \left( 
\frac{ \Delta}{ \omega} \right)^2 }\; ,
\label{65}
\eeq

\no while the energy of the Fermi sea state is given 
by

\beq
E_{FS} = - \frac{\omega^2}{4 d} - \frac{G N}{2}\; .
\label{66}
\eeq

\no The condensation energy is defined as $E_C = E - E_{FS}$,
which in the weak coupling limit, where $\Delta \ll \omega$,
behaves as

\beq
E_C \sim - \frac{ \Delta_{BCS}^2}{2 d}\; ,
\label{67}
\eeq

\no the well known result.

Next, we shall study the shape of the arcs in this model.
The electrostatic field $h(\xi)$ is given by

\begin{equation}
h(\xi) = \rho_0 \; 
\log{ \frac{
(\xi - \omega)( \Delta^2 - \xi \omega +
\sqrt{ (\Delta^2 + \xi^2)(\Delta^2 + \omega^2)} }{
(\xi + \omega)( \Delta^2 + \xi \omega +
\sqrt{ (\Delta^2 + \xi^2)(\Delta^2 + \omega^2)}}} \;.
\label{68}
\end{equation}

\no

\no To simplify matters, Gaudin considered the limit
$ \Delta \ll \omega$,  where eq.~(\ref{68}) becomes

\begin{equation}
h(\xi) = \rho_0 \; 
\log{\frac{
\xi -  \sqrt{ \Delta^2 + \xi^2} }{
\xi +  \sqrt{ \Delta^2 + \xi^2}}}, \qquad  
|\xi| \ll \omega.
\label{69}
\end{equation}

The conserved quantities $\lambda(\ve)$ are given by

\beq
\lambda(\ve) = \frac{g}{4} \log{ \frac{
|(\ve - \omega)( \Delta^2 - \ve \omega +
\sqrt{ (\Delta^2 + \ve^2)(\Delta^2 + \omega^2)}| }{
|(\ve + \omega)( \Delta^2 + \ve \omega +
\sqrt{ (\Delta^2 + \ve^2)(\Delta^2 + \omega^2)}|}}
\quad 
\stackrel{|\ve| \ll \omega}{\longrightarrow} 
\quad
\frac{g}{4} \log{\frac{
|\ve -  \sqrt{ \Delta^2 + \ve^2}| }{
|\ve +  \sqrt{ \Delta^2 + \ve^2}|}} \; .
\label{69a}
\eeq

In the low energy approximation the equation of the equipotential
curves passing through the points $\ve_0 \pm i \Delta$ 
is given by

\beq
{\cal R} \left[ \xi \; 
{\rm log}  \frac{
\xi -  \sqrt{ \Delta^2 + \xi^2} }{
\xi +  \sqrt{ \Delta^2 + \xi^2}}
+ 2  \sqrt{ \Delta^2 + \xi^2}
\right] = 0.
\label{70}
\eeq

\no This curve cuts the real axis inside the interval
$[ - \omega, \omega]$, which contradicts
the assumptions made so far.
 From numerical
studies we can see that for weak couplings
only a fraction of energies $E_\mu$ form
complex conjugated pairs, which in the
limit $N \rightarrow \infty$
organize themselves into arcs, while
the other energies remain real
and located between the lower
energy levels
(see the appendix for
more details on this issue). 
Taking this into account and
using contour arguments,  
it can be shown that 
the BCS eqs.~(\ref{43}), (\ref{44}), and 
(\ref{45}) also hold in cases where $\Gamma$
cuts the energy interval $\Omega$.

In the general case
the curve $\Gamma $ is given by the equation

\begin{equation}
{\cal R} \left[i 
\xi \; \arcsin{ 
\sqrt{ \frac{ 1 + \xi^2/\Delta^2}{ 1 - \xi^2/\omega^2} }}
+ \omega \;  {\rm Argsinh}{
\sqrt{ \frac{ \Delta^2 + \xi^2}{ \omega^2 - \xi^2}} }
 \right] = 0.
\label{79}
\end{equation}

In the limit $\Delta \ll \omega$ eq.~(\ref{79}) turns
into (\ref{70}). We show in the appendix that 
for $\Delta > \omega$
the arc $\Gamma$ does not touch
the interval $[ - \omega, \omega ]$ 
and all the roots $E_\mu$ are complex.
This happens for $g > g_0 =1.13459$
$(\sinh 1/g_0 =1)$.

\section*{VII) Numerical results}

In this section we present the
comparison between the analytic
results derived in section VI,  
and the numerical ones obtained by
solving directly the Richardson's eqs.~(\ref{6})
for the models considered above.

\subsection*{Two-level model}

In this model there are two energy levels, $\pm \ve_1$,
with a  degeneracy $N/2$. Hence 
the Richardson eqs.~(\ref{6}) become

\beq
\frac{1}{G } 
= \frac{N}{2} \left(  \frac{1}{ \varepsilon_1 - E_\nu}
- \frac{1}{ \varepsilon_1 + E_\nu} \right) 
- \sum_{\mu \neq \nu}^{M}  \frac{2}{ E_\mu - E_\nu} \; ,
\label{91}
\eeq

\no with $M= N/2$. The total number of solutions
of this system of equations
is given by the combinatorial number $C^N_M$. 
The solution which corresponds to the ground state
of the BCS Hamiltonian is the one for which 
$E_\mu(G) \rightarrow - \ve_1, \; \forall \mu$ in the limit where
$G \rightarrow 0$. For any non zero  value of $G$, 
all the roots $E_\mu$ form complex conjugate pairs surrounding
the lower energy  $- \ve_1$. 

The reduced coupling constant for this case is taken as $g = GN$.
Our numerical results are presented in units of $\ve_1$, which is 
equivalent to set $\ve_1 = 1$.

%%%%%%%%%%%%%%%%%%%%%%%%%%%%%%%%%%%%%%%%%%%%%%%%%%%%%%%%%%%%%

\begin{figure}[p!]
\begin{center}
\includegraphics[height= 13 cm,angle= -90]{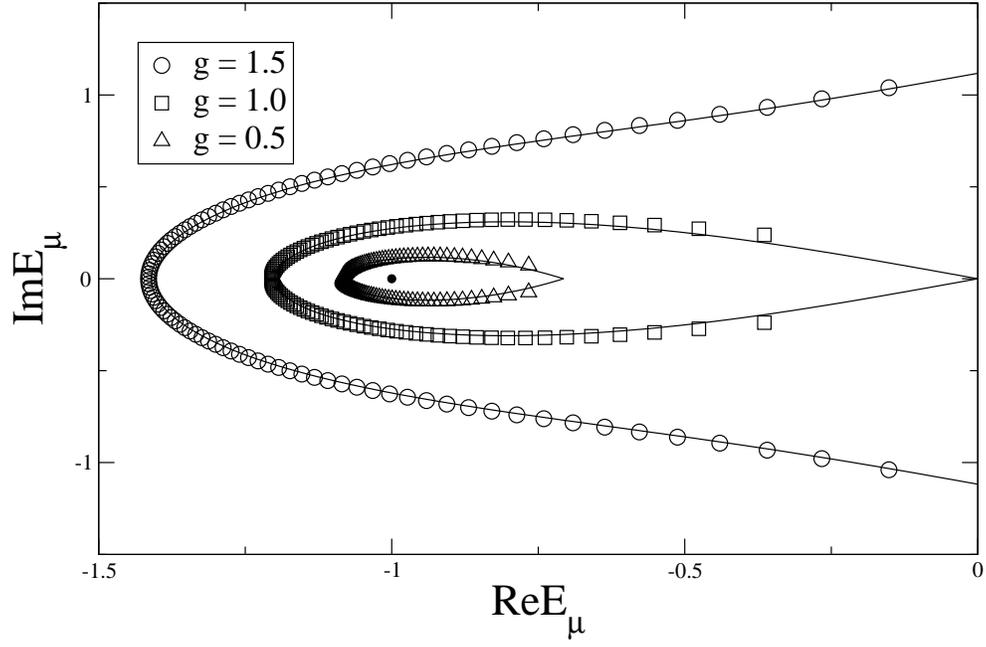}% Here is how to import EPS art
\end{center}
\caption{Plot in the $\xi$-plane of the solutions of  
eq.~(\ref{91}) for M=100 and three values of $g$. The continuous
lines are the theoretical curves (\ref{52}) (with the change of variables
(\ref{47}))  for $g =1.5, 1$; 
and (\ref{58}) for $g=0.5$.
All the energies are in units of $\ve_1$. 
}
\label{fig4}
\end{figure}

\begin{figure}[p!]
\begin{center}
\includegraphics[height= 13 cm,angle= -90]{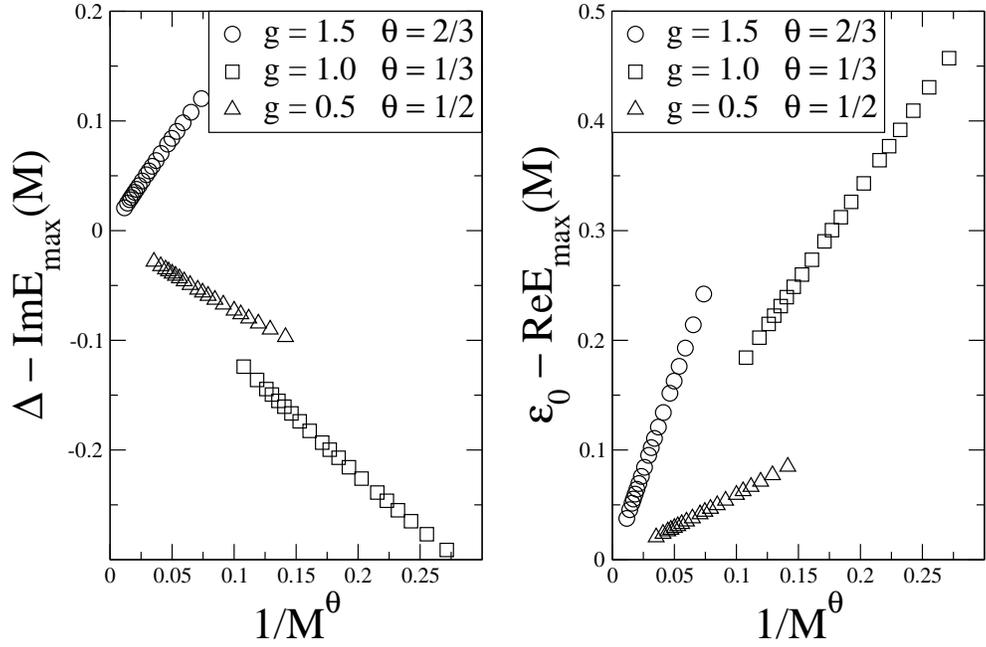}% Here is how to import EPS art
\end{center}
\caption{
Plots of $\Delta - {\rm Im} \; E_{max}(M)$ 
and $\ve_0 - {\rm Re} \; E_{max}(M)$ versus
$M^{- \theta}$, for the two-level model. 
$\ve_0$ and $\Delta$ are given in 
eqs.~(\ref{93}) and (\ref{94}), while 
numerical values  of
$\theta$ are given in  table~1.   
}
\label{fig5}
\end{figure}

%%%%%%%%%%%%%%%%%%%%%%%%%%%%%%%%%%%%%%%%%%%%%%%%%%%%%%%

In fig.~\ref{fig4} we plot the 
distribution of the $M=100$  roots $E_\mu$, for three
different values of the coupling constant $g = G N$, 
together with the analytic curves
derived in section VI. There is an optimal
agreement between the numerical
and analytic results in the three regimes: 
normal ($g < \ve_1$), critical ($g = \ve_1$),
and superconducting ($g > \ve_1$).
As we increase the value of $M$ up to 800 pairs,
the fit of the numerical
data to the analytical curves improves, and in the 
limit $M \rightarrow \infty$ the discrete
roots collapse into the continuous curves. 
In particular, the complex roots $E_{max}(M)$ and  $E_{max}^*(M)$, 
whose real part lies closer
to $\ve_0$, approach their continuum limit value,
namely 

\beq
\lim_{M \rightarrow \infty} E_{max}(M) 
= \ve_0 + i \Delta ,
\label{92}
\eeq

\no where

\qq
\ve_0 & = & \left\{
\begin{array}{cll} 
0 &\qquad & g \geq \ve_1 \\
- \ve_1 \;\sqrt{ 1 - g/\ve_1} &\qquad  & g \leq \ve_1 
\end{array} \right. ,
\label{93} \\
& & \nonumber \\
\Delta & = & \left\{
\begin{array}{cll} 
\sqrt{ g^2 - \ve_1^2} &\qquad & g \geq \ve_1 \\
0 &\qquad & g \leq \ve_1 
\end{array} \right. , 
\label{94}
\qqq

\no in agreement with eqs.~(\ref{50}) and (\ref{59}).

Fig.~\ref{fig5} shows the following 
scaling behaviour  $E_{max}(M)$:
\beq
E_{max}(M) - ( \ve_0 + i \Delta) \sim \frac{1}{M^\theta} \; ,
\label{95}
\eeq
\no where the exponent $\theta$ depends on the regime
of the system. The numerical and analytic results
of $\theta$ are given in table~1:

\begin{center}
\begin{tabular}{|c|c|c|c|}
\hline
$\theta$ & $g = 1.5\; \ve_1 $ &  $g = 1.0\; \ve_1 $ &  $g = 0.5\; \ve_1 $ \\
\hline
analytical & 2/3 &  1/3 & 1/2 \\
numerical $(\Delta)$ & 0.635   & 0.329  & 0.446 \\
numerical $(\ve_0$) & 0.672 & 0.328 & 0.513 \\
\hline
\end{tabular}

\vspace{0.5 cm}
Table 1.- Numerical and analytical values of the exponent
$\theta$ appearing in eq.~(\ref{95}). The numerical values
are extracted from the real and  imaginary parts of
$E_{max}(M)$.
\end{center}

The analytical value of $\theta$ can be derived as 
follows. For $M$ sufficiently large we can write

\beq
E_{max}(M) = ( \ve_0 + i \Delta)  + \delta \xi ,
\label{96}
\eeq

\no where $\delta \xi$ is a small number. 
Using eqs.~(\ref{53a}), (\ref{54}) and (\ref{56}), 
the density of roots $r(E_{max})$ behaves
as

\beq
r(E_{max}) \sim A M  (\delta \xi)^\nu
, \qquad 
\nu = \left\{
\begin{array}{cc}
1/2 & g > \ve_1 \\
2 & g = \ve_1 \\
1 & g < \ve_1 
\end{array} 
\right. ,
\label{97}
\eeq

\no where $A$ is a parameter which only depends
on $g$ and $\ve_1$. 
The magnitude of $\delta \xi$ can be fixed by imposing 
the existence of at least one root in the
interval $( \ve_0 + i \Delta,  \ve_0 + i \Delta +
\delta \xi$), namely

\beq
1 = r(E_{max}) \; |\delta \xi| 
\sim A M \; |\delta \xi|^{\nu + 1}
\quad \longrightarrow \quad
\delta \xi \sim \frac{1}{M^{ \frac{1}{1+\nu}}} \; ,
\label{98}
\eeq

\no which is the desired result (\ref{95})
with $\theta = 1/(1 + \nu)$. 

%%%%%%%%%%%%%%%%%%%%%%%%%%%%%%%%%%%%%%%%%%%%%%%%%%%%%%%%%%

\begin{figure}[p!]
\begin{center}
\includegraphics[height= 13 cm,angle= -90]{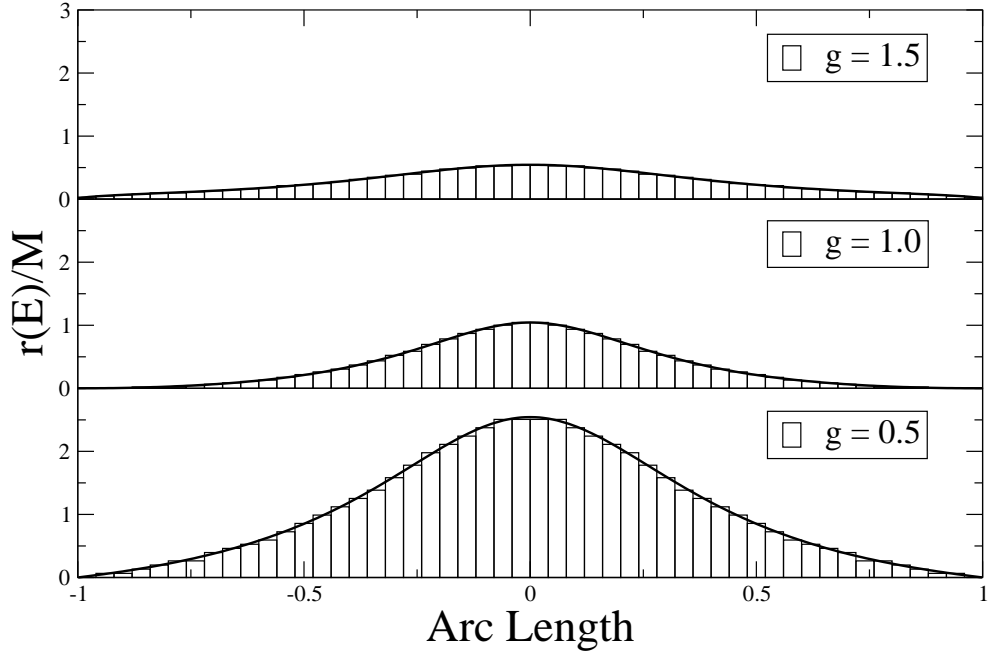}% Here is how to import EPS art
\end{center}
\caption{
Density of roots $r(E)/M$ as a function of the arc length.
The continuous curves give the analytic result while the
step like curves are the numerical results obtained for
M= 800 and three values of $g$. The total
length of the curve has been divided into 50 parts.
The  height of each part is the fraction of 
roots $E_\mu$ per unit length $\delta E$.  
}
\label{fig6}
\end{figure}

\begin{figure}[p!]
\begin{center}
\includegraphics[height= 13 cm,angle= -90]{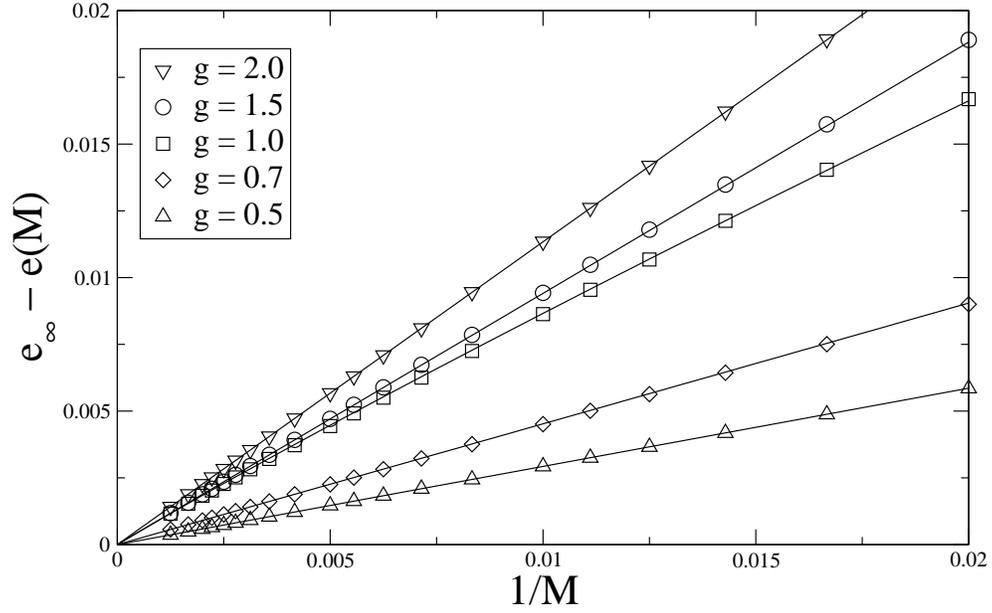}
\end{center}
\caption{Plot of $e_\infty - e(M)$ versus $1/M$ 
for the two-level-model and 
 different values of $g$. The continuous lines 
are given by eqs.~(\ref{99})
for $g  \neq \ve_1$, and eq.~(\ref{99a}) for 
$g= \ve_1$. 
}
\label{fig7}
\end{figure}

%%%%%%%%%%%%%%%%%%%%%%%%%%%%%%%%%%%%%%%%%%%%%%%%%%%%%%%%%%%%

As we have seen in fig.~\ref{fig4}, the roots $E_\mu$
lie  on the analytic curves to a very good degree
of approximation. A further
check of the analytic solution is to compare
the density of roots $r(\xi)$, which is given by
$r(E) = |h(E)|/\pi$  for $g \geq \ve_1$
(see~eq.~(\ref{37})), and by 
$r(\xi) = |s(\xi)|$ for $g \leq \ve_1$
(see~eq.~(\ref{54})). In fig.~\ref{fig6} we plot
the fraction of roots per unit length $r(E)/M$ as a 
function of the arc length $s \in  [-1,1]$, where
the $s = \pm 1$ corresponds to the end points of the
arc $\Gamma$. The agreement is extremely accurate
for $M=800$.

The value of the conserved quantities obtained in eq.~(\ref{53a})
also agree with the numerical ones in the large $N$ limit, 
in the three different regimes.

Finally, it is interesting to study the 
behaviour of the energy as a funtion of the
number of pairs $M$. The leading term is given
by eq.~(\ref{53}) for $g  \geq \ve_1$ and 
by eq.~(\ref{62}) for $g \leq \ve_1$. 
In fig.~\ref{fig7} we plot 
$e_\infty - e(M)$ versus $1/M$, 
where $e(M) = E(M)/M$ is the energy per pair. 
The linear behaviour of the numerical data means that 
the next leading  correction to $E(M)$ is a constant term, 
$E_1$, whose value is given by the analytic expression

\beq
-E_1/M = e_\infty - e(M) =
\left\{ 
\begin{array}{lcl}
 (g - \frac{1}{2} \Delta)/M & \quad & g \geq \ve_1 \\
(\ve_1 - |\ve_0|)/M & \quad  & g \leq \ve_1 
\end{array} \right. ,
\label{99} 
\eeq

\no where $\Delta $ and $\ve_0$ 
 are  given by eqs.~(\ref{50}) and (\ref{59})
respectively. The formula for $g > \ve_1$
has been derived by Richardson in ref.~[\onlinecite{R-limit}], 
studying the next to leading order of the
electrostatic model. The formula for $g < \ve_1$
is an educated guess which needs a proof.
At $g = \ve_1$, the numerical data show a large
deviation, for small $M$,  
from the formula (\ref{99}), namely
$\ve_1/M$. There are theoretical reasons
to believe that the scaling behaviour of
$E_1/M$ at $g = \ve_1$ , is given by the following formula
(see fig.~\ref{fig7}):

\beq
-E_1/M = \ve_1 \left(
\frac{1}{M} - \frac{A}{M^{4/3}} ,
\right),  \qquad A = 0.62258,
\label{99a} 
\eeq

\no which would imply that the next-to-next leading correction of 
the total energy goes as $M^{-1/3}$, rather than
$M^{-1}$. In fact, the formula given by Richardson \cite{R-limit} 
for the $O(M^{-1})$ energy correction, which is given by  
\mbox{$E_2 = - g (3 g - \Delta)( g - \Delta)/(4 N \Delta^2)$},
breaks down at $g=\ve_1$ where the gap vanishes. The derivation
of eq.~(\ref{99a}) shall be given elsewhere.

\subsection*{Equally spaced model}

This model is defined by $N$ non degenerate
energy levels $\ve_j = d (2 j - N - 1),\; j=1, \dots,N$,
where $d=\omega/N$ is the single particle level spacing.
The first numerical study of this model, using the 
exact solution, was done by Richardson in reference~\cite{R-roots}, 
where he introduced new set of variables, 
related to the roots $E_\mu$, 
in order to handle the singularities arising 
at particular values of the coupling constant $g$ (see below). 
This work was revised recently in connection to ultrasmall
metallic grains in [\onlinecite{vDB1,vDB2}].

The ground state of the system, at half-filling $M=N/2$, 
is the one for which $\lim_{g \rightarrow 0} 
E_\mu(g) \rightarrow \ve_\mu \; (\mu=1, \dots, N/2)$.

The reduced coupling constant is given by $g = G/d = GN/\omega$
in this section.  And all our numerical results are presented in
units of $\omega$, which is equivalent to set $\omega = 1$.

%%%%%%%%%%%%%%%%%%%%%%%%%%%%%%%%%%%%%%%%%%%%%%%%%%%%%%%%%%%%%%

\begin{figure}[t!]
\begin{center}
\includegraphics[width= 10 cm]{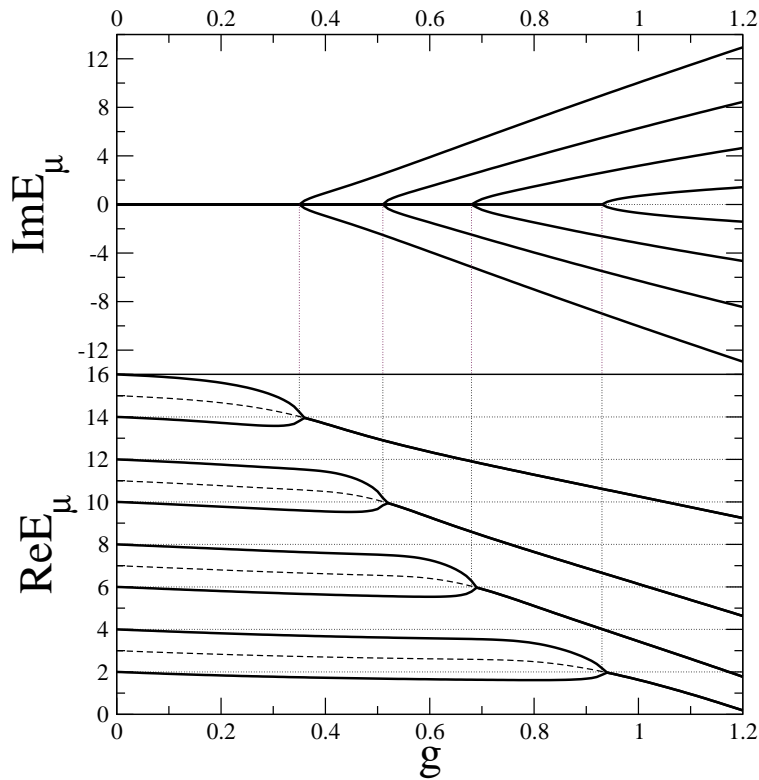}
\end{center}
\caption{Evolution of the real and imaginary parts
of $E_\mu(g)$, in units of $d = \omega/N$, 
for the equally spaced model with $M=N/2=8$, 
as a function of the coupling
constant $g$. For convenience the energy levels
are choosen in this figure as $\ve_j = 2 j$.
This figure confirms numerically 
the behaviour found in reference [\onlinecite{R-roots}].
}
\label{fig8}
\end{figure}

%%%%%%%%%%%%%%%%%%%%%%%%%%%%%%%%%%%%%%%%%%%%%%%%%%%%%%%%%%%%%

In fig.~\ref{fig8} we plot the real and imaginary 
parts of $E_\mu$, for
a system with $M=N/2=8$ pairs. For small values of $g$, all
the pair energies are real and below the corresponding 
energy levels at $g=0$. At a certain value $g_1$,
the two closest pairs to the Fermi energy, namely $E_{M}$
and $E_{M-1}$, coincide at $\ve_{M-1}$ and then, for larger
values of $g> g_1$ they form a complex conjugate pair, while
the rest of the pair energies remain real. 
At higher critical values of $g$ the same phenomena
happens for the other energies, until all of them
become complex. At the critical values of $g$ the Richardson 
eqs.~(\ref{6}) are singular, but the singularities 
can be resolved by the change
of variables introduced in [\onlinecite{R-roots}].

In fig.~\ref{fig9} we show the distribution of roots
for $M=100$ pairs, together with the theoretical 
curves obtained in section VI, for three different values
of $g$. For $g=0.5$ and $1$ there is a fraction of roots
which become complex, falling into the arcs $\Gamma$ described 
by eq.~(\ref{79}), while the real roots extend from 
the cutting point of the arc $\Gamma$ with the real axis
down to the energy $- \omega$. For $g=1.5$ all the roots
are complex and fall into the theoretical curve
(\ref{79}). The numerical data are consistent with the
value $g_0 = 1.13459$ above which all the energies become
complex.

%%%%%%%%%%%%%%%%%%%%%%%%%%%%%%%%%%%%%%%%%%%%%%%%%%%%%%%%

\begin{figure}[p!]
\begin{center}
\includegraphics[height= 13 cm,angle= -90]{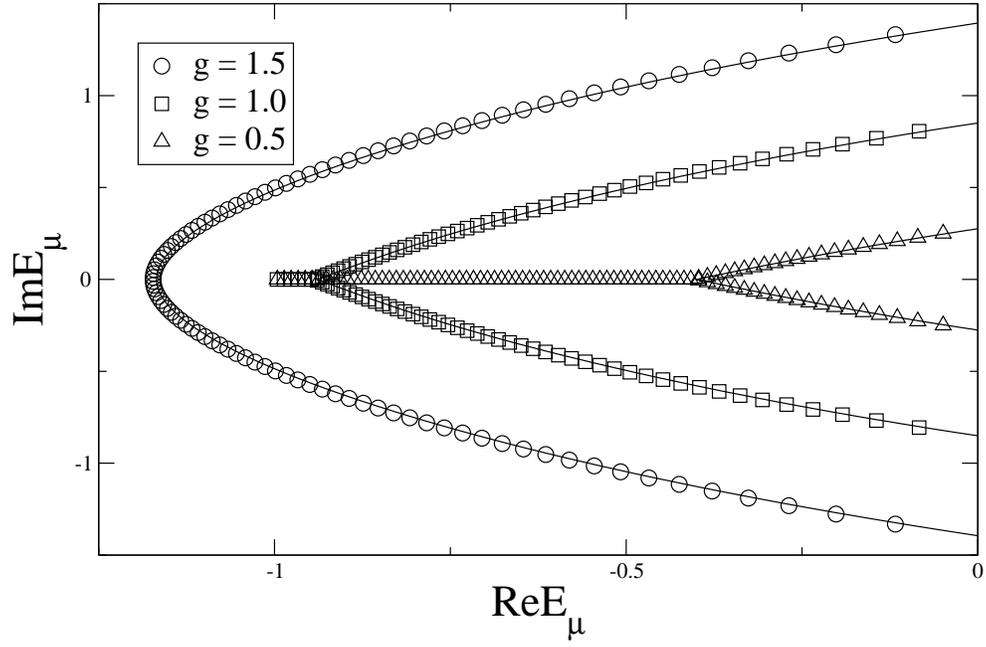}
\end{center}
\caption{Plot of the roots $E_\mu$ for the equally spaced model
in the complex $\xi$-plane. 
The discrete symbols denote the numerical values for $M=100$. 
The continuous lines 
are given by eq.~(\ref{79}). All the energies are in units
of $\omega$.
}
\label{fig9}
\end{figure}

\begin{figure}[p!]
\begin{center}
\includegraphics[height= 13 cm,angle= -90]{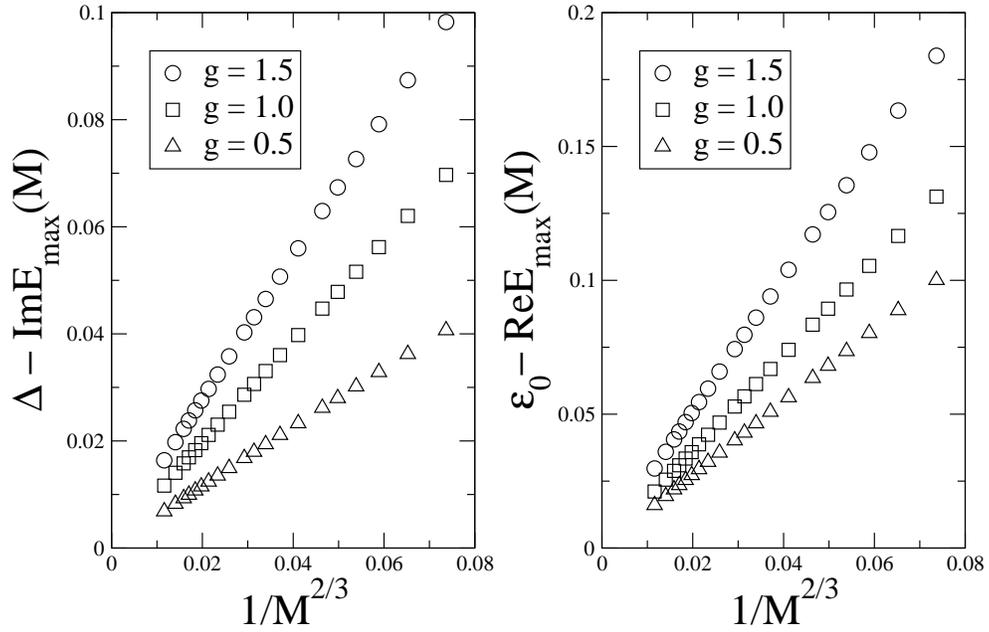}
\end{center}
\caption{Plots of $\Delta - {\rm Im} \; E_{max}(M)$ 
and $\ve_0 - {\rm Re} \; E_{max}(M)$ versus
$M^{- 2/3}$, for the equally spaced model. 
}
\label{fig10}
\end{figure}

%%%%%%%%%%%%%%%%%%%%%%%%%%%%%%%%%%%%%%%%%%%%%%%%%%%%%%%%

As in the two-level case, the root $E_{max}(M)$
closest to $\ve_0 + i \Delta$ satisfy the scaling law
(\ref{95}) with an exponent  $\theta=2/3$. This result
can be proved along the same lines as for
the two-level case. In table~2  we give the 
comparison between the theoretical and numerical
values of $\theta$, and in fig.~\ref{10} we show
the scaling behaviour for three values of $g$.

\begin{center}
\begin{tabular}{|c|c|c|c|}
\hline
$\theta$ & $g =1.5 $ &  $g = 1 $ &  $g =0.5 $ \\
\hline
analytical & 2/3 &  2/3 & 2/3 \\
numerical $(\Delta)$ & 0.646   & 0.645  & 0.642 \\
numerical $(\ve_0$) & 0.658 & 0.659 & 0.661 \\
\hline
\end{tabular}

\vspace{0.5 cm}
Table 2.- Numerical and analytical values of the exponent
$\theta$ appearing in eq.~(\ref{95}) for the equally spaced
model. The notation is the same as in table~1.
\end{center}

%%%%%%%%%%%%%%%%%%%%%%%%%%%%%%%%%%%%%%%%%%%%%%%%%%%%%%%%%%%%%

\begin{figure}[t!]
\begin{center}
\includegraphics[height= 13 cm,angle= -90]{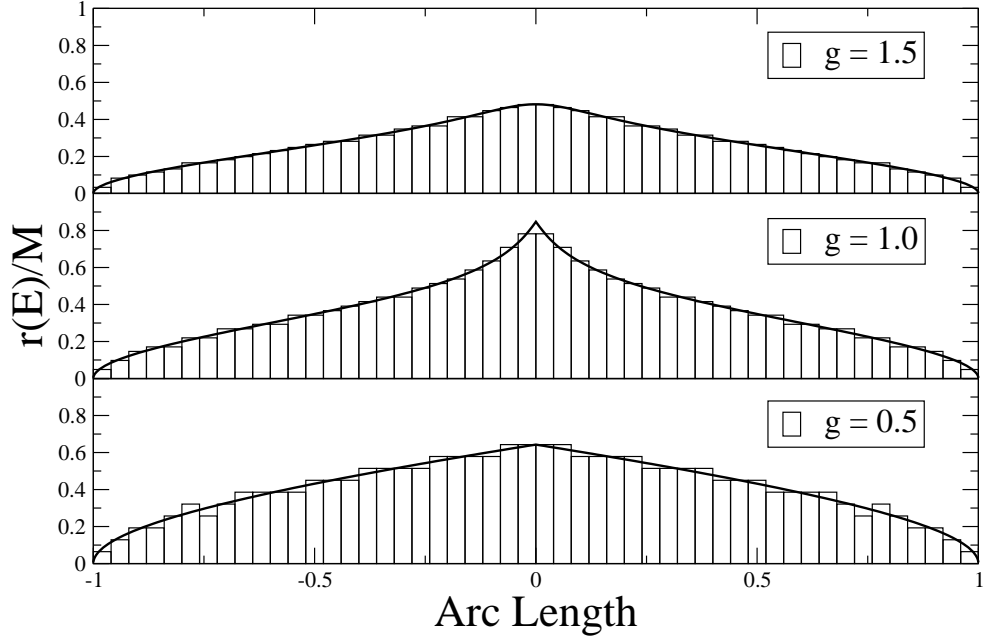}
\end{center}
\caption{Density profile of the complex roots
for the equally spaced model with $M=800$. Notations are the same
as in fig.~\ref{fig6}. 
}
\label{fig11}
\end{figure}

\begin{figure}[p!]
\begin{center}
\includegraphics[height= 13 cm,angle= -90]{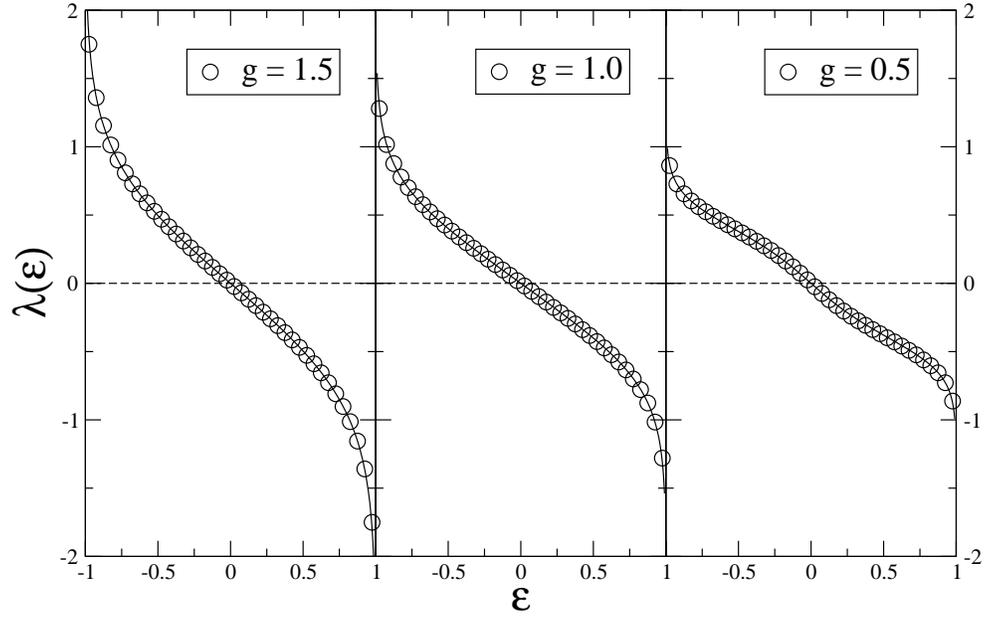}
\end{center}
\caption{Conserved quantities in the region $\Omega$ 
in the equally spaced model
for $M=20$ and different values of $g$. The continous
lines are given by eq.~(\ref{69a}).
}
\label{fig11a}
\end{figure}

%%%%%%%%%%%%%%%%%%%%%%%%%%%%%%%%%%%%%%%%%%%%%%%%%%%%%%%%%%%%

In fig.~\ref{fig11} we give the fraction of roots per
unit length $r(E)/M$ for $M=800$ and three values of $g$.
Again, the numerical data fit accurately the theoretical
result $r(E)/M = |h(E)|/\pi$, where $h(E)$ is given 
by eq.~(\ref{68}). 

A further check is presented in fig.~\ref{fig11a}, where
the distribution of the conserved quantities $\lambda(\ve_i)$
for $M = 20$ is presented, together with their continuous
limit curves given in eq.~(\ref{69a}). A very good
agreement is achieved even  for such a small number of pairs.

%%%%%%%%%%%%%%%%%%%%%%%%%%%%%%%%%%%%%%%%%%%%%%%%%%%%%%%%%%%%

\begin{figure}[p!]
\begin{center}
\includegraphics[height= 13 cm,angle= -90]{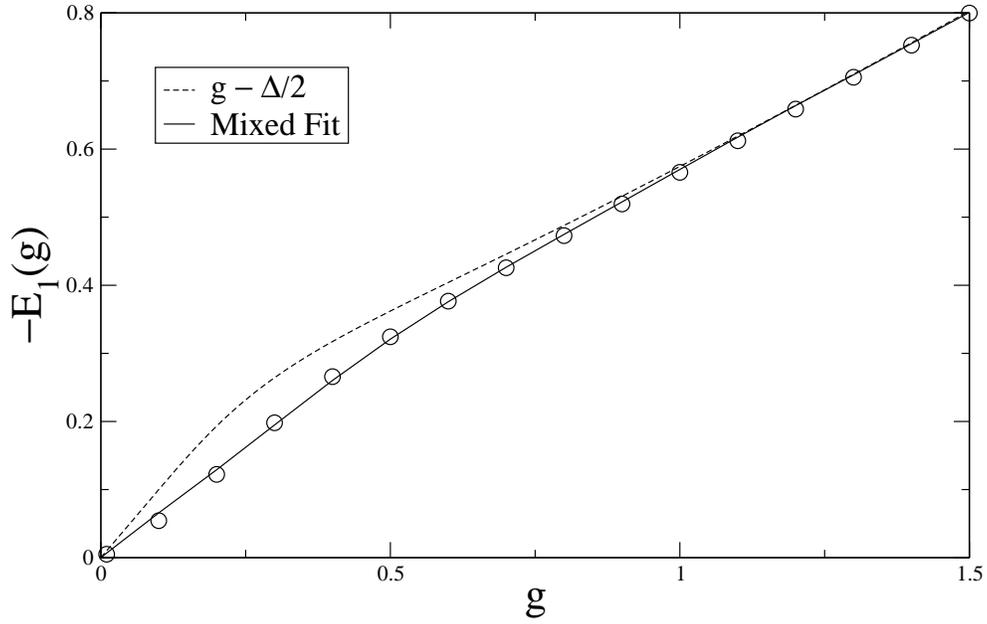}
\end{center}
\caption{
Plot of $-E_1(g)$ versus $g$ 
for the equally spaced level model. The fits of the numerical
data are given by eqs.~(\ref{100}) and (\ref{101}) in units
of $\omega$.  
}
\label{fig12}
\end{figure}

%%%%%%%%%%%%%%%%%%%%%%%%%%%%%%%%%%%%%%%%%%%%%%%%%%%%%%%%%%%%

We have finally studied the condensation energy
$E_C/M$ as a function of $M$, obtaining the behaviour
\mbox{$E_C(M) = e_\infty M + E_1$}, where $E_1(g)$ is plotted in 
fig.~\ref{fig12}. An analytic expression for $E_1(g)$ is not 
known \cite{R-limit}. In order to understand the numerical results
we have made two fits. The first one is based on the result
obtained for the two-level case, namely

\beq
-E_1(g) = g \omega - \frac{\Delta}{2} \; ,
\label{100}
\eeq

\no where $\Delta$ is given by eq.~(\ref{64}). 
Fig.~\ref{fig12} shows that (\ref{100}) is a good
fit for large values of $g$, where all the roots
are complex, forming a single arc. 
For $g \lesssim g_0 = 1.13459$ there is a fraction 
$M_{complex}/M$ of complex roots, and a fraction
$M_{real}/M$ of real roots. This suggests
the following mixed fit:

\qq
-E_1(g)/\omega & = & \left(g - \frac{\Delta}{2 \omega} \right) 
\frac{M_{complex}}{M}
+ (A + B g)\; g \;  \frac{M_{real}}{M} \; , \nonumber \\
& & \label{101} \\
A & = & 0.6735, \qquad B \ =\ -0.1729 ,
\nonumber
\qqq

\no which describes pretty well $E_1(g)$ in the whole
range of $g's$. In the weak coupling regime there have been
proposed several formulas for the finite size corrections
of the condensation energy $E_C$ using the DMRG method
\cite{DMRG} and the exact solution \cite{Sch}. In the latter
reference it is shown that for couplings $g > g* =1/ln N$, 
the condensation energy $E_C$ can be splitted in two contributions, 
one due to the complex energy pairs ( the BCS term) and
another one due to the real roots ( perturbative term). 
It would be interesting to clarify the relation between
formula (\ref{101}) and eq.(24) in reference \cite{Sch}.

\section*{VIII) Conclusions and prospects}

In this paper we have reviewed and completed
Gaudin's continuum limit of the exactly solvable BCS model, 
and compared its predictions with the
numerical solutions describing the ground state of 
the two-level  and the
equally spaced models. 
 We have confirmed previously known
results  \cite{R-roots,random,vDB1,vDB2}, 
and obtained new ones, which we list below:

\begin{itemize}

\item Analytic and numerical 
determination of the curves $\Gamma$, 
where the roots $E_\mu$ of the Richardson equations lie,
together with their density along $\Gamma$. 
For the equally spaced model the 
curves $\Gamma$ are given 
for arbitrary values of the coupling
constant $g$, and for the two-level model 
in the normal regime, i.e. $g < \ve_1$, 
the analytic curves $\Gamma$ differ from those in
[\onlinecite{G-book}]. 

\item Analytic expression of 
the eigenvalues $\lambda_j$ 
of the conserved quantities found by Cambiaggio,
Rivas, and Saraceno \cite{CRS}, which fit
accurately the numerical data, even for a few levels.

\item Study of the scaling properties of the 
ground state energy $E(M)$ 
with the number of pairs $M$ which, in the  
large $M$ limit, behaves generically 
as $E(M) = e_\infty M +
E_1 + O(1/M)$. For the two-level model, in the superconducting
regime, we find an agreement with the analytic
result for $E_1$  [\onlinecite{R-limit}], while in the
normal regime we have guessed an analytic 
expression for $E_1$, 
which fits perfectly the numerical data. In the
critical case there are large deviations in $E(M)$,
which can be explained  by a  term  $O(M^{-1/3})$. 
In the equally spaced model
the condensation energy also receives a constant
contribution, $E_1$,  which is fitted with a ``phenomenological''
formula which combines the effects  of the complex
and real roots.

\item Scaling behaviour of the roots $E_{max}(M)$
which lie closest to the end points of the curves $\Gamma$.
This scaling can be characterized by a critical exponent
$\theta$, whose analytic values fit well the numerical ones.
In the superconducting regimes (i.e. $\forall g > 0$ for  
the equally spaced-model and $g > \ve_1$ for the
two-level model) we find $\theta =2/3$, while in the
normal (critical)
regimes of the two-level model we find 
$\theta = 1/2$ ($ 1/3$). In this manner the exponent $\theta$
characterizes the nature of the ground state. An interesting 
question is wether $\theta$ shows up in physical observables.

\item Generalization of Gaudin's conformal mapping for the
equally spaced model using elliptic functions, which degenerate
into trigonometric ones in the limit $\Delta \ll \omega$. This suggest
the existence of a rich mathematical structure.

\end{itemize}

\no
Finally, we shall mention some problems which are worth
to investigate along the lines suggested by the present work:

\begin{itemize}

\item Study of the excited states and the finite size
effects in a more geometrical
way, completing  the work initiated by Richardson in
[\onlinecite{R-limit}].

\item Quantum dots at  
finite temperature is 
another difficult problem, which has been treated combining
several techniques depending on the temperature regime 
\cite{Finite-T,FFM,FF}.
Is it possible to find a thermodynamic Bethe ansatz for this system?

\item Study of solutions of Richardson equations
with several arcs, 
i.e.\ $ K > 1$ in the notation of section V.
For the standard BCS model
they  must  describe very high excited
states formed by separate condensates
in interaction. This case may be
relevant to systems such as arrays of superconducting 
grains or quantum dots.
 From a mathematical point of view, the cases
with $K >1$ seem to be related to the 
theory of hyperelliptic curves and higher
genus Riemann surfaces, which may shed some light
on this physical problem.

\item Generalization of Gaudin's equations to
Richardson models based on  arbitrary
Lie groups ${\cal G}$ \cite{AFS}. The standard 
case corresponds to the choice ${\cal G} = SU(2)$.  
The electrostatic model for generic  ${\cal G}$
has charges labelled by vectors 
of dimension ${\rm rank}\; {\cal G}$. 
It is naturally to expect 
that the roots $E_{\mu}^a \; (a=1, \dots, 
{\rm rank}\;  {\cal G})$
should fall into several arcs labelled by the index 
$a=1, \dots, {\rm rank} \;  {\cal G}$.

\end{itemize}

\section*{Acknowledgments}

We thank Jo\~ao Lopes dos Santos, Rodolfo Cuerno, 
Miguel Angel Mart\'{\i}n-Delgado, and Javier Rodr\'{\i}guez-Laguna
for computational help, and Didina
Serban for the reference [\onlinecite{G-book}]. 
This work has been supported by the Spanish grants 
BFM2000-1320-C02-01 (JMR and GS), and 
BFM2000-1320-C02-02 (JD).

%%%%%%%%%%%%%%%%%%%%%%%%%%%%%%%%%%%%%%%%%%%%%%%%%%%%%%%%%%%%%%%%%%%%%%%
%%%%%%%%%%%%%%%%%%%%     APPENDICES       %%%%%%%%%%%%%%%%%%%%%%%%%%%%%
%%%%%%%%%%%%%%%%%%%%%%%%%%%%%%%%%%%%%%%%%%%%%%%%%%%%%%%%%%%%%%%%%%%%%%%

%\pagebreak

\pagebreak

\appendix

\section*{Appendix:  Conformal mapping for the equally spaced system}

In this appendix
we study the analytic structure underlying the
shapes of the curves defined by eqs.~(\ref{70}) and (\ref{79}).

\subsection*{i) Case  $\Delta \ll \omega$}

In this  limit it is convenient to make 
a conformal
mapping \cite{G-book}
from the upper half-plane ${\rm Im} \xi > 0$,
with a cut in the segment $(0, i \Delta)$,
into the half-strip $\lambda > 0$, $- \pi/2 < \mu < \pi/2$, 
\beq
\xi = i \Delta \;  \cosh( \lambda + i \mu).
\label{71} 
\eeq
\no In the plane $u= \lambda + i \mu$ the curve (\ref{70})
becomes
\beq
\lambda \; \tanh \lambda + \mu \; \cot \mu = 0.
\label{72}
\eeq

%%%%%%%%%%%%%%%%%%%%%%%%%%%%%%%%%%%%%%%%%%%%%%%%%%%%%%%%

\begin{figure}[p]
\begin{center}
\includegraphics[height= 12 cm,angle= -90]{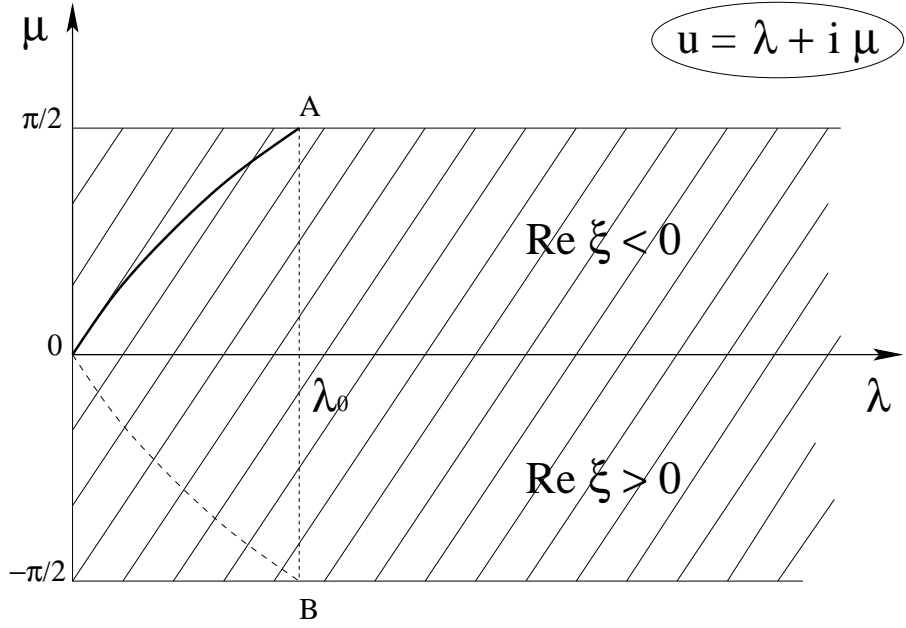}
\end{center}
\caption{Half-strip in the $u=\lambda + i \mu$ plane
that is conformally equivalent to the upper half $\xi$-plane, 
with a cut in the interval $[0, \ve_0 + i \Delta]$. The arcs
${\bf OA}$ and ${\bf OB}$ are given by eq.~(\ref{72}). This
figure is borrowed from reference \cite{G-book}. 
}
\label{fig1}
\end{figure}

\begin{figure}[p]
\begin{center}
\includegraphics[height= 12 cm,angle= -90]{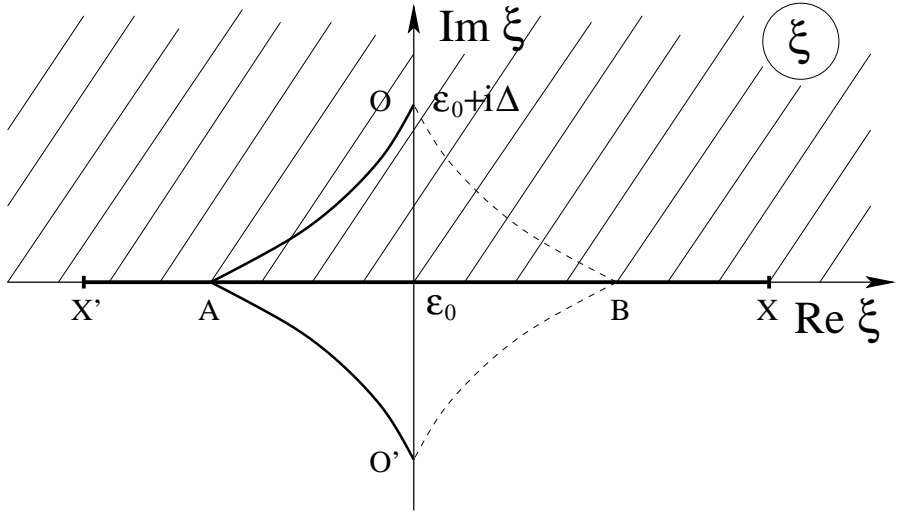}
\end{center}
\caption{Location of the roots $E_\mu$ in the $\xi$-plane. 
The complex roots condense in the arc ${\bf OAO'}$, given
by eq.~(\ref{70}), while the real ones are 
distributed in the segment ${\bf X'A}$, with  a real
root between every two energy levels. The point ${\bf A}$
is given by eq.~(\ref{73}).  This
figure is borrowed from reference \cite{G-book}. 
}
\label{fig2}
\end{figure}

%%%%%%%%%%%%%%%%%%%%%%%%%%%%%%%%%%%%%%%%%%%%%%%%%%%%%%%%

\no
Figures \ref{fig1} and \ref{fig2} show, in the 
$u$ and $\xi$ planes, the arcs defined by
eqs.~(\ref{72}) and (\ref{70}). The point   
${\bf O}$ ($\xi= \ve_0 + i \Delta)$ 
is mapped into the origin ($u=0$), 
where the field $h$ 
vanishes. In fact $h(\xi)$, as a function of $u$,
is simply $h = - 2 \rho_0 u$. 
The point ${\bf O'}$ ($\xi= \ve_0 - i \Delta)$ 
is the mirror reflection of ${\bf O}$. The arc
${\bf OA}$ is the one that corresponds to the
actual numerical solution for ${\rm Im} \xi >0$,
while its mirror image, 
 ${\bf O'A}$, is the one for ${\rm Im} \xi < 0$.  
We thus see that the whole arc
${\bf OAO'}$ cuts the real axis at the point 
${\bf A}$ with an energy
\beq
\ve_A = \ve_0 - \Delta \; \sinh{\lambda_0}, \qquad
\lambda_0 \; \tanh{\lambda_0} = 1. 
\label{73}
\eeq
\no 
These results are expected from numerical
studies, which show that for weak couplings
only a fraction of energies $E_\mu$ form
complex conjugated pairs, which in the
limit $N \rightarrow \infty$
organize themselves into arcs, while
the other energies remain real
and located between the lower
energy levels ( see fig.~\ref{fig9}). We can check this result
computing the number of real and complex
roots $E_\mu$. From eq.~(\ref{73})
we deduce that  
the real roots
occupy the interval $[ - \omega, - \Delta 
{\rm sinh} \lambda_0]$, with a density
which is $2 \rho_0$.  Hence 
the total number, $M_{\rm real}$  of real
roots is 
\beq
M_{\rm real} = 2 \rho_0 
( \omega - \Delta \; \sinh{\lambda_0} ).
\label{74}
\eeq
\no Since the total number of roots is
$M = M_{\rm real} + M_{\rm complex}= N/2 = 2 \rho_0 \omega$,
we deduce that 
\beq
M_{\rm complex} = 2 \Delta  \rho_0 \; \sinh{\lambda_0}.
\label{75}
\eeq
\no This result agrees with the one
 obtained integrating 
the field $h(\xi)$  along the curve $L$
surrounding $\Gamma$
(recall eq.~(\ref{44})). This integration
is more easily done in the $u$-strip
\beq
M_{\rm complex} = \int_L h(\xi) \; d \xi =
4 \times  \int_{0}^{\lambda_0 + i \pi/2} (- 2 \rho_0 u)
(i \Delta  \sinh{u} \; du),
 \label{76}
\eeq
\no where the factor 4 comes from the contribution
of the up and down, and left and right pieces
of the contour $L$. Using contour arguments,  
it can be shown that 
the BCS eqs.~(\ref{43}), (\ref{44}), and 
(\ref{45}) also hold in cases where $\Gamma$
cuts the energy interval $\Omega$. 
As a side comment, we observe that
$M_{\rm complex}= \frac{\Delta}{2 d} \;
{\rm sinh}  \lambda_0$, agrees
with the heuristic argument stating that 
there are roughly
$\Delta_{BCS}/d$ energy levels
around the Fermi level that are strongly affected
by the pairing interactions \cite{T}. See also
reference \cite{Sch} for an approximate equation
giving the number of complex pairs ( eq. (49)).

\subsection*{ii) Generic case}

In order to compare with the numerical
results presented in the  section VII, 
it is convenient not to make the approximation
$\Delta \ll \omega$. The appropiate
change of variables
that generalizes (\ref{71}) is given by 
\beq
\xi = 
\frac{ i \Delta \; \cosh{u}}{\displaystyle{
\sqrt{ 1 -  \left(  \frac{\Delta}{\omega}\; \sinh{u} \right)^2}} }\;,
\label{77} 
\eeq
\no in terms of which $h = - 2 \rho_0 u$. 
The integral $\int h(\xi) d \xi$ can be more
easily done in the $u$ plane, yielding for
the equation of the curve $\Gamma$
\begin{equation}
{\cal R} \left[
\frac{ i\;  u \; \cosh{u}}{\displaystyle{
\sqrt{ 1 -  \left(  \frac{\Delta}{\omega}\; \sinh{u} \right)^2}} } 
- \frac{i \;  \omega}{\Delta} \; \arcsin\left( \frac{\Delta}{\omega}
\; \sinh{u} \right) \right] = 0,
\label{78}
\end{equation}
\no which in the $\xi$-plane gives eq.~(\ref{79}), namely
\begin{equation}
{\cal R} \left[i 
\xi \; \arcsin{ 
\sqrt{ \frac{ 1 + \xi^2/\Delta^2}{ 1 - \xi^2/\omega^2} }}
+ \omega \;  {\rm Argsinh}{
\sqrt{ \frac{ \Delta^2 + \xi^2}{ \omega^2 - \xi^2}} }
 \right] = 0.
\label{79bis}
\end{equation}
\no In the limit $\Delta \ll \omega$ this equation turns
into eq.~(\ref{70}). We can ask when this curve 
cuts the interval $[- \omega, \omega]$. A convenient
parametrization of the cutting point is (recall eq.~(\ref{73}))
\beq
\xi_A = - \frac{ \Delta \; \sinh{\lambda_0}}{\displaystyle
\sqrt{ 1 + \left( \frac{ \Delta}{\omega} \; 
\cosh{\lambda_0} \right)^2} } \;.
\label{80}
\eeq
\no Plugging (\ref{80})
into (\ref{79bis}) we deduce the equation for
$\lambda_0$,
\beq
\frac{ \lambda_0 \; \sinh{\lambda_0}}{\displaystyle
\sqrt{ 1 + \left( \frac{ \Delta}{\omega} \; 
\cosh{\lambda_0} \right)^2}}
= \frac{\omega}{\Delta} \; {\rm Argsinh}
\left( \frac{ \Delta}{\omega} \; 
\cosh{\lambda_0} \right) ,
\label{81}
\eeq
\no which only has real solutions provided
$\Delta \leq \omega$. When $\Delta > \omega$
the arc $\Gamma$ does not touch
the interval $[ - \omega, \omega ]$ 
and all the roots $E_\mu$ are complex.
This happens for $g > g_0 =1.13459$. This value
of $g_0$ is obtained solving the equation 
$\sinh{1/g_0} =1$.

\medskip

To complete this appendix we shall show that
eq.~(\ref{77}) yields a conformal
mapping that generalizes the one found 
by Gaudin for generic values of  $\Delta/\omega$.
To do so, let us first perform the change
of variables
\beq 
u = i \left( \frac{\pi}{2} - \phi \right),
\label{82}
\eeq
\no and introduce the parameters 
\beq
k = \frac{\Delta}{\sqrt{\Delta^2 + \omega^2}},
\qquad
k'=  \frac{1}{\sqrt{\Delta^2 + \omega^2}} =
\sqrt{1 - k^2} \, ,
\label{83}
\eeq
\no in terms of which (\ref{77}) becomes
\beq
\xi = 
\frac{ i \Delta k' \; \sin{\phi}}{
\sqrt{ 1 -  k^2 \; \sin^2{\phi}} } \;. 
\label{84} 
\eeq
\no Using the elliptic functions with  modulus 
$k$, and conjugate modulus $k'$, 
we define a new variable $z$ through \cite{WW}
\qq
\sin{\phi} & = & {\rm sn} (z,k), \nonumber \\
\cos{\phi} & = & {\rm cn} (z,k),
\label{85} \\
\sqrt{ 1 - k^2 \; \sin^2{\phi}} & = & {\rm dn} (z,k).
\nonumber 
\qqq
\no The relation between $z$ and $\phi$ is given 
by the equation
\beq
z(k, \phi) = \int_0^\phi \frac{ d \psi}{ \sqrt{ 1 - k^2 \; 
\sin^2{\psi} }} \; ,
\label{86}
\eeq
\no while the inverse function is denoted as
$\phi = {\rm am} \;(z,k)$. 
The relation between $\xi$ and $z$ reads
\beq
\xi = i \Delta k' \; \frac{{\rm sn}(z,k)}{{\rm dn}(z,k)} \; .
\label{87}
\eeq
\no Finally, using the relation  
${\rm sn}(K + v,k)/{\rm dn}(K+v,k) = {\rm cn}(v,k)/k'$,
where $K$ is the half-period of the elliptic integrals,
eq.~(\ref{87}) turns into
\beq
\xi = i \Delta  \; {\rm cn}(v,k),
\label{88}
\eeq
\no where $z= K+v$. The relation between the variables
$u$ and $v$ is given by
\beq
i u = - \frac{ \pi}{2} + {\rm am}(K + v, k) 
= \frac{\pi v}{2 K} + 
\sum_{n=1}^\infty \;
\frac{2 q^n}{ n (1 + q^{2 n} } \; 
\sin{\frac{ n \pi ( K + v)}{2 K}} \; ,
\label{89}
\eeq
\no where $q= e^{ - \pi K'/K}$. In the limit 
$\Delta/\omega \ll 1$ we get 
$k \rightarrow 0$, $K \rightarrow \pi/2$, 
and $q \rightarrow 0$, which implies that $i u \rightarrow
v$ and eq.~(\ref{88}) reduces to  (\ref{71}). 

%%%%%%%%%%%%%%%%%%%%%%%%%%%%%%%%%%%%%%%%%%%%%%%%%%%%%%%%%%%%%%%%%%%%%%

\begin{figure}[t!]
\begin{center}
\includegraphics[height= 12 cm,angle= -90]{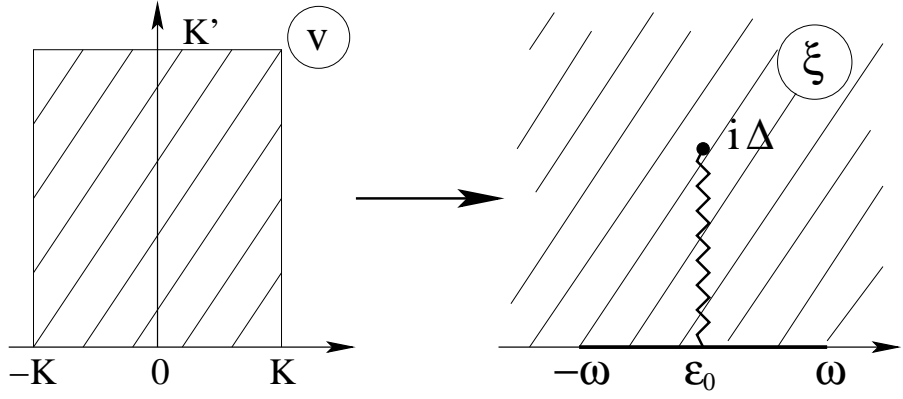}
\end{center}
\caption{Graphical representation of the conformal
map (\ref{88}), with the correspondences given in table 3.
}
\label{fig3}
\end{figure}
   
%%%%%%%%%%%%%%%%%%%%%%%%%%%%%%%%%%%%%%%%%%%%%%%%%%%%%%%%%%%%%%%%%%%%%%

Eq.~(\ref{88}) gives a conformal map of
the rectangle $-K < {\rm Re} \; v < K, \;  0 < {\rm Im} \; v < K'$
onto the upper half-plane $ {\rm Im} \; \xi > 0$, with a cut
along the linear segment $ 0 < {\rm Im} \; \xi \leq \Delta$
(see reference [\onlinecite{N}] and fig.~\ref{fig3}). 
Some examples of this mapping are given by 
\beq
v = 0, \pm K, \pm K + i K' 
\qquad \longrightarrow \qquad 
\xi = i \Delta, 0, \pm \omega \; ,
\label{90}
\eeq
\no which imply the following identifications
between the sides of the rectangle
in the $v$-plane and the relevant energy
regions in the $\xi$-plane:
\begin{center}
\begin{tabular}{|c|c|c|}
\hline
cut  & $
\begin{array}{c}
{\rm Re} \; \xi = 0 \\
0 < {\rm Im}\; \xi < \Delta
\end{array}$  
 &  $\begin{array}{c} 
- K < {\rm Re} \; v < K \\
{\rm Im} \; v = 0 \end{array}$ \\
\hline
lower band & $
\begin{array}{c}
 - \omega < {\rm Re} \; \xi < 0 \\
{\rm Im} \; \xi = 0 \end{array}$
&  
$\begin{array}{c}
{\rm Re} \; v = - K \\
0 < {\rm Im} \; v < K' 
\end{array} $ 
\\ \hline
upper band & $ 
\begin{array}{c}
0 < {\rm Re} \; \xi < \omega \\
{\rm Im} \; \xi = 0 \end{array}
$  &  
$\begin{array}{c}
{\rm Re} \; v =  K \\
0 < {\rm Im} \; v < K' 
\end{array} $ 
\\ \hline
off band & $ 
\begin{array}{c}
| {\rm Re} \; \xi | > \omega \\
{\rm Im} \; \xi = 0 \end{array}
$  &  
$\begin{array}{c}
- K < {\rm Re} \; v <  K \\
 {\rm Im} \; v = K' 
\end{array} $ 
\\ \hline
\end{tabular}

\vspace{0.5 cm}
Table 3.- Correspondences of the boundary regions
of the conformal map (\ref{88}). 
\end{center}

\end{document}